\documentclass[11pt]{article}
\usepackage{iftex}
\ifPDFTeX
  \usepackage[T1]{fontenc}
  \usepackage[utf8]{inputenc}
  \usepackage{lmodern}
  \usepackage{textcomp}
\else
  \usepackage{fontspec}
\fi
\usepackage[a4paper,margin=1in]{geometry}
\usepackage{amsmath,amssymb}
\usepackage{graphicx}
\usepackage{booktabs}
\usepackage{tabularx}
\usepackage{array}
\usepackage{multirow}
\usepackage{subcaption}
\usepackage{float}
\newcolumntype{Y}{>{\centering\arraybackslash}X}
\usepackage{xcolor}
\usepackage{titlesec}
\usepackage{fancyhdr}
\usepackage{enumitem}
\usepackage{tcolorbox}
\tcbuselibrary{skins,breakable}
\usepackage[numbers,sort&compress]{natbib}
\usepackage[colorlinks=true,linkcolor=faircblue,citecolor=faircblue,urlcolor=faircteal]{hyperref}

\definecolor{faircnavy}{HTML}{0E2A47}
\definecolor{faircblue}{HTML}{2E6DA4}
\definecolor{faircteal}{HTML}{1C7C88}
\definecolor{faircgrey}{HTML}{5A6672}
\definecolor{panel}{HTML}{F4F7FA}
\definecolor{mustred}{HTML}{C0392B}
\definecolor{shouldamber}{HTML}{B9770E}
\definecolor{couldslate}{HTML}{5B6B7A}

\titleformat{\section}{\Large\bfseries\color{faircnavy}}{\thesection.}{0.6em}{}
\titleformat{\subsection}{\large\bfseries\color{faircblue}}{\thesubsection}{0.6em}{}
\titleformat{\subsubsection}{\normalsize\bfseries\color{faircteal}}{}{0em}{}
\titlespacing*{\section}{0pt}{16pt}{6pt}

\newtcolorbox{recbox}[2]{
  enhanced, breakable, colback=panel, colframe=faircblue, boxrule=0.8pt,
  arc=3pt, left=10pt, right=10pt, top=8pt, bottom=8pt,
  title={\textbf{#1}\hfill\mbox{\textbf{#2}}}, coltitle=white, colbacktitle=faircblue,
  fonttitle=\normalsize }

\begin{document}

% =================== TITLE PAGE ===================
\begin{titlepage}
\thispagestyle{empty}
\vspace*{1.2cm}
\begin{flushleft}
{\color{faircblue}\rule{\linewidth}{2.5pt}}\\[10pt]
{\Huge\bfseries\color{faircnavy} FAIR-Compute}\\[10pt]
{\LARGE\bfseries A Roadmap for Fair and Efficient Allocation of\\ Federated Digital Research Infrastructure}\\[14pt]
{\large\color{faircgrey} Final Report to the National Federated Compute Services (NFCS) Flexible Fund}\\[6pt]
{\large\color{faircgrey} Project short code: \textbf{FAIRC}}\\[24pt]
{\color{faircteal}\rule{\linewidth}{1pt}}\\[10pt]
{\large\textbf{Principal Investigator}}\\[3pt]
{\large Professor Konstantinos (Kostas) E. Zachariadis}\\
{\normalsize School of Economics and Finance, Queen Mary University of London}\\[14pt]
{\large\textbf{Project team}}\\[3pt]
\begin{tabular}{@{}ll@{}}
Dr Ahmed Sayed & Co-lead --- Electronic Engineering \& Computer Science, QMUL\\
Professor Dimitris Fotakis & Advisor --- Electrical \& Computer Engineering, NTUA\\
Angeliki Mathioudaki & Researcher (simulation \& model) --- NTUA \& ICCS\\
Wan Shuen Siaw & Researcher (roadmap \& field data) --- Mathematical Sciences, QMUL\\
\end{tabular}
\\[24pt]
{\normalsize\color{faircgrey} Final project report \quad\textbullet\quad Submitted July 2026}
\end{flushleft}
\vfill
\end{titlepage}

\tableofcontents
\newpage

% =================== EXECUTIVE SUMMARY ===================
\section*{Executive Summary}
\addcontentsline{toc}{section}{Executive Summary}

As demand for high-performance computing (HPC), high-throughput computing, and data storage grows, the way scarce compute is \emph{allocated} --- not just how much exists --- has become a decisive factor in the productivity of UK research. FAIR-Compute studies allocation in a federated Digital Research Infrastructure (DRI) as a problem at the intersection of algorithmic game theory, transport economics, and HPC scheduling. Our central observation is simple: once a shared system must decide who runs, when, and under what evidence, its scheduler settings cease to be a purely technical matter and become \textbf{policy}.

The project combined three strands of evidence: a landscape review of UK and international allocation practice (including EuroHPC, WLCG, ACCESS, JASMIN, and DiRAC) supported by a stakeholder survey; a mechanism-design model of allocation under strategic and uncertain user reports; and a simulation study built on the public Fresco/Anvil workload trace and controlled synthetic stress tests. Three results recur across all three strands. First, allocation records measure \emph{occupancy} (resources reserved) rather than \emph{utilisation} (useful work done), so the system cannot currently answer the question UKRI and DSIT most want answered --- whether resources deliver productive value. Second, simple, transparent scheduling heuristics perform close to an offline full-information benchmark once tuned, so the near-term opportunity is to tune and instrument existing schedulers rather than replace them. Third, federation is genuinely valuable but behaves like uncoordinated (``selfish'') routing when left unmanaged: it can minimise \emph{average} delay while quietly concentrating load --- and potentially harm --- on the receiving system. This mirrors a classic effect from road-traffic economics (Braess's paradox), where letting everyone independently pick the fastest route can leave the whole network worse off --- so coordinated placement rules are needed to contain it.

From this evidence we derive \textbf{eight recommendations} for the NFCS roadmap, categorised Must / Should / Could Have and spanning a five-year horizon. The foundation is measurement: a unified data-collection standard (\textbf{FAIRC-1}) and an occupancy-versus-utilisation efficiency KPI (\textbf{FAIRC-2}), both Must-haves on which the others depend. These are followed by practical, low-risk governance improvements --- publishing Slurm tuning guidance (\textbf{FAIRC-4}), a tiered rapid-access route paired with Community Centres of Excellence (\textbf{FAIRC-7}), cross-system benchmarking and exchange rates (\textbf{FAIRC-6}), and ex-post verification of requested-versus-consumed resources (\textbf{FAIRC-8}) --- and two forward-looking pilots: value-aware prioritisation (\textbf{FAIRC-3}) and coordinated governance of federated job mobility (\textbf{FAIRC-5}). If the roadmap adopts one thing first, it should be measurement: almost everything else depends on data that does not yet exist in a consistent form.

% =================== INTRODUCTION ===================
\section{Introduction and Approach}

The UK is moving from a set of independently governed HPC systems towards a federated National Compute ecosystem in which access, platform services, and infrastructure are increasingly disaggregated and coordinated centrally. This transition raises an allocation question that existing, single-institution scheduling heuristics were not designed to answer: how should a heterogeneous, multi-owner federation match a diverse supply of compute to a heterogeneous, strategically-behaving demand, fairly and efficiently, when the information needed to decide is imperfect and privately held?

FAIR-Compute frames this as a \textbf{mechanism-design} problem rather than a purely engineering one. Allocation has two layers that must be considered together: the \emph{scheduler}, which fixes ordering and placement (FIFO, backfilling, aging); and the \emph{policy mechanism}, which shapes incentives and priorities (fairshare, quality-of-service tiers, walltime penalties, credits). Users may misreport their needs, either strategically (to gain priority) or through genuine uncertainty about runtimes, and the resulting behaviour --- most visibly walltime ``padding'' --- is best treated as a permanent feature of the system to be modelled, not a bug to be wished away.

The project was organised around three complementary activities, and the report follows their logic. Section~\ref{sec:method} sets out our \textbf{methodology and evidence base} in detail --- the stakeholder interviews and their follow-up, the demand-side survey, the literature review, and the design of the simulation study --- so that the basis for every later claim is transparent. Section~\ref{sec:landscape} presents the resulting \textbf{landscape review}: the UK and international allocation ecosystem, the economic evidence, and the operational frictions surfaced by our case studies. Section~\ref{sec:model} develops the \textbf{formal model} of allocation under strategic and uncertain reporting that underpins the simulations. Section~\ref{sec:sim} reports the \textbf{simulation study} and its four experiments in full, with methods, results, and figures. Section~\ref{sec:recs} states the eight \textbf{roadmap recommendations} in the NFCS output format, each tied back to this evidence. Section~\ref{sec:future} sets out the \textbf{future-work} agenda the project opens up, and Section~\ref{sec:conclusion} concludes. The appendices collect the technical detail --- evaluation metrics, the workload proxy, the full federation policies and results, the engagement programme, and reproducibility notes --- so that the main narrative stays readable while the underlying work remains fully documented.

% =================== METHODOLOGY ===================
\section{Methodology and Evidence Base}
\label{sec:method}

FAIR-Compute combined three strands of evidence, each chosen to compensate for the blind spots of the others. A \emph{landscape and literature review} established what is known --- from both academic mechanism-design theory and operational practice in the UK and abroad. \emph{Stakeholder engagement} (semi-structured interviews and a demand-side survey) grounded that theory in how allocation actually works, and fails, on live national systems. And a \emph{simulation study} let us run controlled, counterfactual experiments that no single operator could run on a production system. Triangulating across the three is what gives the recommendations in Section~\ref{sec:recs} their weight: each is supported by theory, by operator testimony, and, where possible, by simulation evidence.

\subsection{Stakeholder interviews and follow-up}
The primary qualitative evidence comes from a series of semi-structured interviews conducted between February and June 2026 with operators, directors, and policymakers spanning the breadth of the UK ecosystem. In total we held in-depth sessions with seven organisations: QMUL ITS Research (operators of the Apocrita Tier-3 cluster); the STFC IRIS / IRIS-SCRC allocation team; the JASMIN data-intensive facility (NERC); the DiRAC distributed HPC service for theoretical physics and cosmology; the EPCC at Edinburgh; the UKRI Digital Research Infrastructure (DRI) programme; and the Department for Science, Innovation and Technology (DSIT) public-compute division, including a dedicated session on the AI Research Resource (AIRR). Several stakeholders were consulted more than once as our questions sharpened; Appendix~\ref{app:engagement} summarises the full engagement programme.

Each interview followed a prepared agenda covering user behaviour and demand (estimation accuracy, walltime padding, strategic job-splitting), mechanism and tier design (preemption logic, pricing and cost-recovery, Slurm priority weights), and the transition to federation (identity mapping, cross-charging, data gravity). Crucially, interviews were treated as an \textbf{iterative} process rather than one-off data collection. After each session the findings were written up and \emph{mapped onto the formal model's variables} --- for example, IRIS-SCRC's penalty loop and DiRAC's under-utilisation review thresholds were mapped to the historical usage state that reappears in the model (Section~\ref{sec:model}) and in recommendation FAIRC-8; JASMIN's storage-binding constraint motivated the dual-resource extension noted under FAIRC-2; and the recurring ``occupancy versus productive use'' theme became a through-line of the whole project. Where operators raised open questions --- most importantly how to define ``productive use'' and how to share pseudonymised job logs --- these were carried back into subsequent sessions and into formal data requests, several of which remain in progress and directly motivate FAIRC-1.

Analysis was thematic rather than statistical, which suits a small number of deep expert interviews. Findings were synthesised into recurring themes (occupancy versus utilisation; the administrative bottleneck; batch versus interactive tension; contested fairness; the difficulty of measuring outcomes) and, importantly, \emph{cross-checked across independent sources}: a claim was only treated as robust when it recurred across organisations with different incentives --- for instance, the observation that heuristics ``work fairly well but we don't know how far from optimum'' surfaced independently at an institutional cluster, a national credit-based service, and within DSIT. This triangulation is what licenses the report to generalise from a modest number of interviews.

\paragraph{The data-request process.} A specific, and instructive, strand of the engagement was the attempt to obtain real job-scheduling data to calibrate the simulations. We requested pseudonymised historical job logs (submission and actual runtimes, resource requests, queue and scheduling decisions) and, where possible, anonymised ``reason-for-rejection'' records, and we discussed acceptable sharing terms (pseudonymisation at source, removal of system-identifying labels). Despite constructive engagement, no UK dataset could be released on workable terms within the project window --- a combination of privacy, administrative, and effort constraints that operators themselves flagged as a barrier. This experience is not a footnote but a finding: the absence of a routine, agreed data-sharing pathway is precisely the gap that recommendation FAIRC-1 is designed to close, and it is why the simulation study (Section~\ref{sec:sim}) had to fall back on public US workload data.

\subsection{Demand-side stakeholder survey}
To complement the operator-facing interviews with a demand-side view, we ran an exploratory stakeholder survey with a user-facing focus. The instrument comprised \textbf{16 questions} covering resource-estimation behaviour, allocation and fairness perceptions, scheduling strategies, and federation challenges, and was administered through Microsoft Forms. Data collection ran over a two-month period from \textbf{April to June 2026}. The survey was distributed through a combination of conference QR codes (including at the UKRI NFCS Spring Conference in Bristol), LinkedIn posts, and direct professional networks, allowing respondents to participate electronically.

A total of \textbf{ten responses} ($N=10$) were collected. Respondents comprised system administrators, HPC facility managers, and research software engineers ($n=6$), together with funder, policy, and institutional stakeholders ($n=4$); several reported interacting with more than one class of infrastructure (national supercomputers, university clusters, and cloud HPC). Respondents are therefore best understood as people who support, manage, or represent users --- and who observe user behaviour daily across many accounts --- rather than as a representative sample of end users, and we interpret the responses accordingly. We note two further limitations transparently. First, the survey did \emph{not} capture respondents' Research Council affiliation (e.g.\ EPSRC, STFC, BBSRC), which limits any analysis by funding community. Second, the sample is small, so the findings should be read as \emph{exploratory and qualitative} rather than statistically representative of the wider UK community. Responses were analysed thematically; the resulting themes --- persistent walltime overestimation, strategic buffering, the absence of any single agreed definition of fairness, and the predominantly organisational (rather than technical) nature of federation barriers --- are reported in Section~\ref{sec:frictions} and corroborate the interview evidence.

\subsection{Literature review}
In parallel we conducted a structured review of the state of the art, organised into four areas that together frame the allocation problem: the \emph{engineering status quo} (Slurm-style priority scheduling, fairshare, and backfilling); \emph{markets with money} for heterogeneous data-centre resources; \emph{coordination mechanisms} and the price of anarchy in congested, self-interested systems; and \emph{mechanism design without money}, including artificial-currency and priority-based approaches. This review both positioned FAIR-Compute against existing work and supplied the theoretical scaffolding --- congestion games, welfare benchmarks, and incentive compatibility --- used in the model and simulations that follow.

\subsection{Simulation study: rationale}
Finally, because no operator can ethically run counterfactual scheduling experiments on a live national system, and because pseudonymised UK job logs could not be obtained on workable terms within the project window, we built a simulation framework on public workload data. Its design, data handling, and four experiments are documented in full in Section~\ref{sec:sim}; here we simply flag the guiding principle. We deliberately separate \emph{real-trace replay} (which validates realism but is low-contention) from \emph{controlled synthetic stress tests} (which create the contention under which allocation policy actually bites), and we label every result accordingly, so that the reader always knows whether a number reflects observed behaviour or a sandbox counterfactual.

% =================== LANDSCAPE / ROADMAP ===================
\section{The UK and International Research-Computing Landscape}
\label{sec:landscape}

\subsection{Infrastructure tiers and federation models}
UK research computing is organised in tiers: international-scale facilities (Tier~0, e.g.\ Isambard-AI, EuroHPC, PRACE); the national flagship service (Tier~1, ARCHER2, which ends in 2026 and is to be replaced by the new Edinburgh-based Next National Supercomputing Service, NNSS, from 2027); regional and specialist hubs (Tier~2, e.g.\ Bede, Baskerville, Cirrus); and institutional systems (Tier~3, e.g.\ QMUL's Apocrita)~\citep{epsrc2019impact}. Researchers typically progress up this ladder as their computational needs grow, but each tier is separately governed, which fragments access.

Four organisational models coordinate such systems: \emph{independent} (local control, but fragmented); \emph{grid} (shared resources, but complex middleware, exemplified by the Worldwide LHC Computing Grid, WLCG~\citep{wlcg2020mou}); \emph{cloud} (elastic, but with cost and interconnect limits for tightly-coupled HPC); and \emph{federated} (coordination with retained institutional autonomy). Federation is the UK's chosen direction. International practice offers concrete mechanisms: the EuroHPC Federation Platform provides federated single sign-on (via the MyAccessID identity service) together with cross-system allocation and resource management through the Waldur platform~\citep{eurohpc_efp,waldur}; Open OnDemand lowers onboarding friction through a common web portal~\citep{hudak2018openondemand}; and the US ACCESS programme operates a credit-and-exchange model in which portable allocation credits are converted between heterogeneous resources using benchmark-derived exchange rates~\citep{access2026resources,access2026projecttypes}.

The four models trade integration against autonomy in different ways (Table~\ref{tab:models}), and each leaves a characteristic residue --- fragmentation, middleware overhead, cost, or governance complexity. Federation is attractive precisely because it aims to keep local control while sharing the services (identity, allocation, software) that most benefit from coordination.

\begin{table}[htbp]\centering
\caption{Organisational models for coordinating HPC systems.}
\label{tab:models}
\small
\begin{tabularx}{\linewidth}{@{}lXX@{}}
\toprule
\textbf{Model} & \textbf{Strength} & \textbf{Weakness} \\
\midrule
Independent & Full local control & Fragmentation; multiple accounts \\
Grid & Resource sharing across sites & Middleware complexity; steep workflows \\
Cloud & Elastic, on-demand provisioning & Cost and interconnect limits for tightly-coupled HPC \\
Federated & Coordination with retained autonomy & Governance and cross-charging complexity \\
\bottomrule
\end{tabularx}
\end{table}

Recent UKRI proposals point to a \textbf{disaggregated stack} --- access, platform services, and compute as independent layers behind one ``single pane of glass'' portal --- coordinated by \textbf{Community Centres of Excellence} (CCEs) that broker allocations across federated systems and provide training and research-software-engineering (RSE) support, alongside tiered access with rapid ``Seedcorn / Bright Idea'' routes~\citep{ukri_townhall2026,etp4hpc2025federation}.

\subsection{Occupancy, utilisation, and the economic case}
A recurring theme across every operational interview is the gap between \textbf{occupancy} --- resources reserved or blocked for a job --- and \textbf{utilisation} --- resources actually performing useful computation. A system can show high occupancy and low utilisation (reserved-but-idle capacity), which is pure opportunity cost. The natural efficiency measure is therefore utilisation divided by occupancy, yet current accounting almost universally charges for occupancy, which rewards over-requesting and walltime padding.

Published economic evaluations show that HPC investment delivers strong returns, but they are scarce and methodologically inconsistent (Table~\ref{tab:bcr}). EPSRC's evaluation of its federated Tier-1/Tier-2 programme reports a benefit--cost ratio (BCR) of 6.5--19.5~\citep{epsrc2019impact}. The Met Office programme reports a 9:1 cost--benefit ratio \emph{against the do-nothing option}, based on a net present social value of \pounds13.74bn (with a 25\% optimism bias) on a \pounds1.2bn investment; because that ratio is not a simple benefit-to-investment quotient, it is not directly comparable with the EPSRC figure~\citep{metoffice_ao2022}. Industry estimates warrant particular caution: for NVIDIA's privately-funded Cambridge-1, Frontier Economics estimated an economic value of roughly \pounds600M ($\sim$\$831M) over ten years on a \$100M investment~\citep{frontier2021cambridge1}, and other cited figures vary. Even at face value, these numbers do not establish that federation yields higher returns than independent provision --- only that HPC broadly pays off and that comparable, standardised ROI evidence is missing. That measurement gap is itself an argument for recommendations FAIRC-1 and FAIRC-2.

\begin{table}[htbp]\centering
\caption{Published benefit--cost ratios for public HPC infrastructure. The ratios derive from different evaluation methodologies and should \emph{not} be read as a direct comparison between infrastructure models (see text); the purpose of the table is to show that published studies consistently find substantial economic value from HPC investment across different contexts. We report the headline ratios and their basis rather than derived investment/benefit figures, which vary by source. NVIDIA Cambridge-1 is discussed in the text as an industry estimate.}
\label{tab:bcr}
\small
\begin{tabularx}{\linewidth}{@{}lcX@{}}
\toprule
\textbf{Case study} & \textbf{BCR} & \textbf{Basis and source} \\
\midrule
EPSRC Tier-1 \& Tier-2 HPC (federated) & 6.5--19.5\,:\,1 & Ten-year return on EPSRC's HPC investment; independent evaluation by London Economics for EPSRC~\citep{epsrc2019impact}. \\[3pt]
Met Office supercomputing (independent) & 9\,:\,1 & Ratio against the do-nothing option; net present social value \pounds13.74bn (25\% optimism bias) on a \pounds1.2bn investment~\citep{metoffice_ao2022}. \\
\bottomrule
\end{tabularx}
\end{table}

\subsection{Where allocation breaks down}
\label{sec:frictions}
Our interviews and case studies surface a consistent set of structural frictions that any federated allocation mechanism must confront.

\paragraph{The administrative bottleneck.} Across facilities, manual, peer-review-based allocation --- not hardware --- is the binding scaling constraint. IRIS-SCRC and DiRAC both operate multi-month allocation cycles: a call opens, a technical team reviews requested hardware, a committee decides, and letters issue months later. Reviews are labour-intensive --- operators described allocation reviews consuming on the order of 60--100 person-hours, and DiRAC assessing roughly 67 applications per round under a ``double jeopardy'' process that judges scientific merit \emph{and} computational justification independently. As AI and interdisciplinary demand grows, this human review becomes the scaling limit, producing slow cycles, high overhead, and limited responsiveness. This is the direct motivation for a tiered rapid-access route (FAIRC-7).

\paragraph{Occupancy versus productive use.} Every operator distinguished, in their own terms, between blocking a resource and using it. Standard accounting charges for \emph{occupancy} (core-hours reserved) rather than \emph{utilisation} (work delivered), which rewards conservative over-requesting and walltime padding. Worse, allocation units correlate poorly with scientific output: as one director put it, the amount of science obtained from a given allocation is ``poorly correlated with the allocation units being used,'' and knowing the relationship in advance requires prior domain knowledge that most users lack. Interactive and software-defined environments sharpen the problem, because they must stay ``on'' to preserve state and therefore block resources even when idle.

\paragraph{Batch versus interactive, and the technology shift.} Legacy batch processing guarantees fair access and system stability but is poorly suited to code development, cluster-optimisation research, and cybersecurity work. Software-defined environments (containers, VMs) let users treat a cluster as a ``larger laptop'' through a single pane of glass, lowering the barrier to entry --- but hardware is currently partitioned statically between batch and interactive modes, and making that boundary \emph{dynamic} is itself an unsolved allocation problem.

\paragraph{Contested fairness.} There is no single agreed definition of ``fair.'' Stakeholders variously invoke scientific merit, proportionality to funding or contribution, equal access, and alignment with national strategic priorities. Fairness is therefore a multi-dimensional, contested objective, not a single quantity to optimise --- a point that recurs in the survey findings below and shapes the model in Section~\ref{sec:model}.

\paragraph{Contrasting facility models.} The frictions manifest differently depending on the binding constraint, as our two most detailed case studies show (Table~\ref{tab:jasmin-dirac}). \textbf{JASMIN} (NERC) is a data-intensive facility of roughly 55{,}000 CPU cores and 60--80\,PB of tiered storage, where \emph{storage occupancy}, not compute, is the scarce resource; compute on its Lotus cluster runs under a light-touch FairShare model per consortium, and spare capacity is exposed to the wider federation as a residual claimant. \textbf{DiRAC}, serving theoretical physics, astronomy, and cosmology across four hardware-software co-designed sites, is compute-bound: it allocates CPU/GPU-hour \emph{credits} through peer review, expires them quarterly on a ``use-it-or-lose-it'' basis, links priority to historical use, and triggers follow-up review below 50\% (small) or 80\% (large) utilisation. These are not competing designs but rational responses to different scarcities --- which is precisely why a federation needs mechanisms flexible enough to span both.

\begin{table}[htbp]\centering
\caption{Two contrasting national facility models, from our operator interviews. The binding constraint drives the allocation mechanism.}
\label{tab:jasmin-dirac}
\small
\begin{tabularx}{\linewidth}{@{}lXX@{}}
\toprule
 & \textbf{JASMIN (data-intensive)} & \textbf{DiRAC (compute-intensive)} \\
\midrule
Primary constraint & Storage occupancy (60--80\,PB) & Compute (CPU/GPU-hours) \\
Compute allocation & Light-touch FairShare per consortium & Credit-based, peer-reviewed \\
Temporal policy & Historical-need consortium quotas & Quarterly ``use-it-or-lose-it'' credits \\
Behavioural lever & Minimal; ``no monopoly'' principle & Under-use review (50\%/80\% thresholds) \\
Federation posture & Residual claimant (shares spare compute) & Specialised, co-designed per site \\
\bottomrule
\end{tabularx}
\end{table}

\subsection{Demand-side findings}
The demand-side stakeholder survey (Section~\ref{sec:method}; $N=10$) --- answered largely by those who support and represent users rather than by end users themselves --- corroborates the operator picture from the other side of the interface. Three themes stand out. On \textbf{resource estimation}, respondents reported systematic walltime overestimation with inconsistent accuracy, confirming a persistent information asymmetry: users know more about their workloads than the scheduler does, and they hedge that uncertainty through conservative buffering and backfill-oriented submission. On \textbf{fairness}, responses revealed no consensus definition --- merit, funding-proportionality, equal access, and national-priority alignment were all invoked, reinforcing that fairness is contested rather than given. On \textbf{federation}, the barriers respondents raised were as much organisational (governance, cross-charging, data movement) as technical.

These demand-side frictions fall hardest on particular groups. Early-career researchers without an established PI or local RSE support, and non-traditional HPC disciplines (biology, medicine, the social sciences) that increasingly need large-scale compute but lack established workflows, face the steepest structural barriers: complex onboarding, multiple accounts and proposals, command-line and batch-scheduler expertise, and fragmented identity and software stacks. The consequence is both underused national infrastructure and higher barriers to entry for exactly the interdisciplinary communities the federation is meant to broaden --- the demand-side case for CCEs and a rapid-access tier (FAIRC-7).

\subsection{International comparison and lessons}
Other national and international systems offer concrete design lessons. The \textbf{Worldwide LHC Computing Grid} demonstrates cost-sharing by pledge and, critically, a maturing benchmarking discipline: its move from the HS06 benchmark to \emph{HEPScore} --- containerised real experiment workloads, reproducible to better than 1\% across x86 and ARM --- is the template for the exchange-rate framework we recommend (FAIRC-6)~\citep{giordano2023hepscore}. The US \textbf{ACCESS} programme shows that portable, benchmark-derived credits can let researchers move computational value across heterogeneous machines without re-applying, decoupling allocation approval from machine selection. Further afield, \textbf{China} and \textbf{Japan} illustrate the two poles of the curiosity-versus-mission tension: China's centralised, state-led model prioritises strategic mission workloads, whereas Japan's federated HPCI balances mission and curiosity-driven access through a peer-reviewed, quota-plus-priority system~\citep{qian2018china,caict2022index,hpci2024overview}. The UK's disaggregated, CCE-coordinated direction is a deliberate attempt to occupy the middle ground --- integration with retained autonomy --- and the mechanisms in this report are aimed squarely at making that middle ground work.

\subsection{Sustainability and carbon}
Allocation efficiency is also an energy and carbon question. Idle-but-reserved capacity (high occupancy, low utilisation) draws power without producing science, and operating costs are dominated by energy: DiRAC, for instance, reports roughly \pounds2M of a \pounds5.5M annual operating budget as electricity. Raising realised utilisation per unit of reserved capacity --- the KPI proposed in FAIRC-2 --- therefore lowers energy and carbon per unit of research output, and curbing walltime over-reservation (FAIRC-8) has the same effect. We do not quantify carbon in tonnes of CO\textsubscript{2} equivalent (tCO\textsubscript{2}e) in this study, which is a limitation; we recommend that the efficiency KPI be reported alongside energy consumption so that the carbon impact of allocation policy can be tracked consistently across the federation.

% =================== MODEL ===================
\section{A Model of HPC Allocation}
\label{sec:model}

This section sets out the formal model that gives the project its analytical spine. It has three parts: a description of the system and how a production scheduler actually allocates resources; a decision-theoretic account of how strategic, uncertain users choose what to report; and an offline welfare benchmark against which practical rules can be measured. The simulations in Section~\ref{sec:sim} implement the descriptive scheduler and the offline benchmark directly; the strategic-utility analysis (Section~\ref{sec:model-utility}) frames the mechanism-design questions that motivate recommendations FAIRC-3 and FAIRC-8 and is, in part, a direction for future work. Throughout, the guiding stance --- shared with our operator interviewees --- is that strategic behaviour such as walltime padding is a permanent feature of the system to be modelled, not a defect to be assumed away.

\subsection{System, jobs, and the descriptive scheduler}
A cluster provides a per-time-unit capacity $L=(L_1,\dots,L_d)$ across $d$ resource types (CPU cores, GPUs, memory, \dots). Each job $j$ is a tuple $\beta_j=(a_j,r_j,l'_j)$ of arrival time $a_j$, resource requirement $r_j\in\mathbb{N}^d$, and \emph{reported} walltime $l'_j$; its \emph{actual} runtime $l_j$ is realised only during execution, and a job is terminated if $l_j>l'_j$. An allocation assigns a start time $s_j\ge a_j$; completion is $c_j=s_j+l_j$; feasibility requires $\sum_j x_j(t)\le L$ for all $t$.

Production schedulers such as Slurm~\citep{slurm_documentation} do not solve an explicit welfare objective; they rank eligible jobs by a weighted priority score
\begin{equation}
\mathrm{Priority}_j(t) = w_{\text{age}}\,\mathrm{Age}_j(t) + w_{\text{fs}}\,\mathrm{Fair}_{i(j)}(t) + w_{\text{qos}}\,\mathrm{QoS}_j + w_{\text{size}}\,\mathrm{Size}_j + w_{\text{part}}\,\mathrm{Part}_j + \cdots,
\end{equation}
where $\mathrm{Age}_j(t)=t-a_j$ and the fairshare term decreases with a user's recent usage through a public history state $H_i(t)$ that decays exponentially~\citep{yalim2020toward}. The non-negative weights $w_\cdot$ are set by administrators, not users. Among eligible jobs $\mathcal{E}(t)=\{j\text{ waiting}: r_j\le L_{\text{left}}(t)\}$ the highest-priority job is scheduled first, with smaller jobs backfilled when they do not delay it.

\subsection{Strategic reports and user utility}
\label{sec:model-utility}
Users are typically uncertain about runtimes at submission, knowing only a distribution $l_j\sim F_j$ over possible execution times. The reported walltime $l'_j$ therefore functions as \emph{self-selected insurance}: requesting a larger walltime reduces the probability that the job is killed before completion, but typically increases queue delay (a larger reservation is harder to place and to backfill) and may expose the user to ex-post penalties. The value a user derives from a completed job degrades with delay. Writing the waiting time $W_j=s_j-a_j$, we model completion value as $v_j(W_j)=V_j\,f_j(W_j)$ for a non-increasing urgency function $f_j$; the linear special case $v_j(W_j)=V_j-\kappa_jW_j$ captures a constant marginal disutility of waiting $\kappa_j$ (we reserve $c_j$ for completion times). Crucially, $W_j$ itself depends on $l'_j$ through the scheduler, so a larger safety buffer carries an implicit upfront cost.

Adding a mechanism-imposed ex-post penalty $h_j(l'_j,l_j,H_{i(j)})$ that depends on the reporting gap and on the user's public history, expected utility is
\begin{equation}
\mathbb{E}[U_j(l'_j)] = \mathbb{E}\!\left[\,V_j - \kappa_j W_j - h_j(l'_j,l_j,H_{i(j)})\,\right],
\end{equation}
so the reported walltime is a genuine strategic choice, trading termination risk against queue delay and penalty exposure. The specific termination rules (kill-at-walltime versus preempt-and-requeue) and penalty forms --- an overestimation charge, or a fairshare adjustment that routes the cost through the history state $H_{i(j)}$, as IRIS-SCRC and DiRAC do in practice --- are set out in Appendix~\ref{app:strategic}. The strategic-reporting mechanism itself is \emph{not} exercised in the present simulations; Experiment~3 empirically illustrates the related system-wide externalities that arise from runtime information reporting, and the potential loss of scientific value under value-blind scheduling. This part of the model is the mechanism-design foundation for recommendation FAIRC-8 and for the research programme in Section~\ref{sec:future}.

\subsection{Offline welfare benchmark}
\label{sec:model-benchmark}
To measure how far practical rules sit from an ideal, we use an offline benchmark that schedules with full information. With binary variables $y_{j,t}$ (job $j$ runs at $t$) and $z_{j,t}$ (job $j$ completes at $t$), subject to arrival ($y_{j,t}=0$ for $t<a_j$), capacity ($\sum_j r_{j,k}y_{j,t}\le L_k$), completion ($\sum_{\tau=a_j}^{t}y_{j,\tau}\ge l_j z_{j,t}$), and single-completion ($\sum_t z_{j,t}=1$) constraints, the benchmark maximises total value net of waiting cost,
\begin{equation}
\max \sum_{j\in J}\bigl(V_j - \kappa_j W_j\bigr), \qquad W_j=\sum_{t}(t-a_j-l_j)\,z_{j,t}.
\end{equation}
This weighted-flow-time / value objective is the reference against which FIFO, Slurm-like, and tuned heuristics are compared in Section~\ref{sec:sim}. One practical caveat applies to how the benchmark should be read. Although it is offline and full-information \emph{by construction}, solving it exactly on the workload windows used for comparison is computationally demanding, and the solver did not always certify global optimality, instead terminating with a small but non-zero optimality gap. The offline results reported in Section~\ref{sec:sim} should therefore be understood as the best feasible solution found within the available computational limits --- with an objective value close to the solver's best bound --- rather than as a proven global optimum. This distinction also helps explain why a heuristic can occasionally appear to match or outperform the offline benchmark on individual reported metrics.

% =================== SIMULATIONS ===================
\section{Simulation Study}
\label{sec:sim}

\subsection{Data, synthetic generation, and framework}
\begin{figure}[htbp]\centering
\begin{subfigure}{0.48\linewidth}\centering
\includegraphics[width=\linewidth]{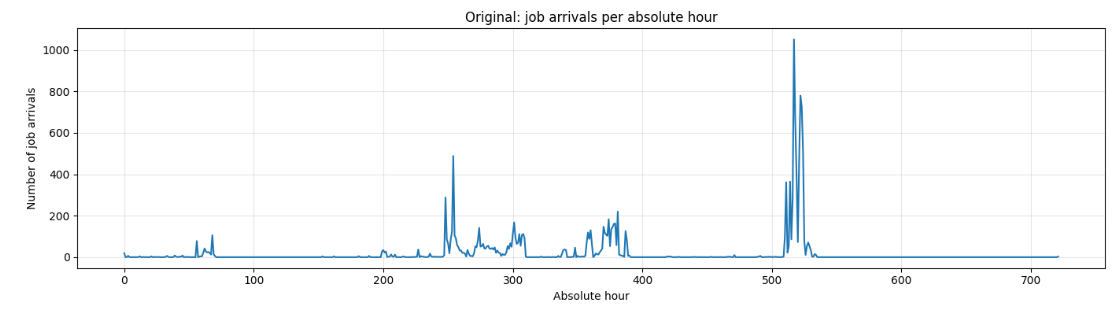}
\caption{Arrivals per hour of day}
\end{subfigure}\hfill
\begin{subfigure}{0.48\linewidth}\centering
\includegraphics[width=\linewidth]{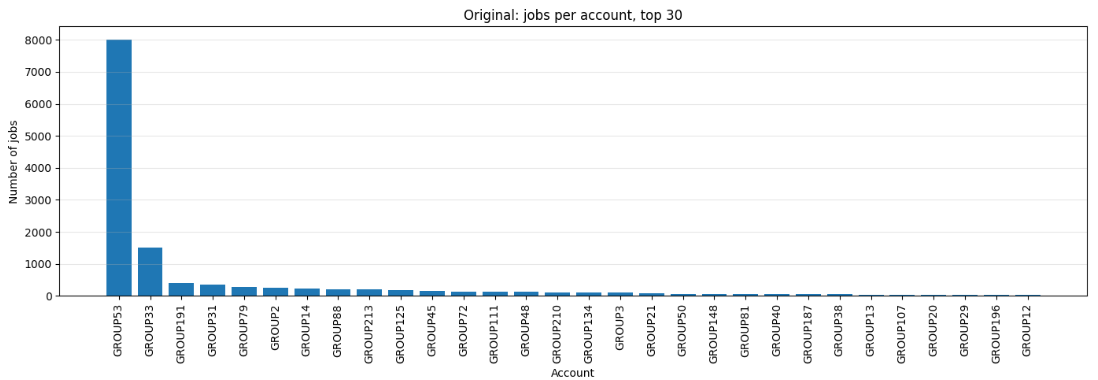}
\caption{Jobs per account (top 30)}
\end{subfigure}
\caption{Descriptive statistics of the Anvil workload: bursty temporal demand (left) and a heavy-tailed distribution of jobs across accounts (right).}
\label{fig:datastats}
\end{figure}

\paragraph{Data.} No internal UK HPC trace was available on workable terms within the project window, so the study uses the public Fresco dataset~\citep{mckerracher2025fresco,fresco_data} --- roughly 20.9 million jobs across 75 months from three US academic systems --- focusing on Purdue's \emph{Anvil} cluster~\citep{song2022anvil}, the newest and, crucially, the only Slurm-based system, over June 2022--May 2023. Anvil records exactly what a scheduling simulation needs: submit, start and end times, requested resources and walltime, queue/partition, account identifiers, and completion status. Because Fresco is distributed as monthly files, account-level history can be discontinuous at month boundaries; we therefore treat each month as a self-contained trace for descriptive and low-pressure analysis, and carry state explicitly where history matters. Since the Fresco trace does not expose the true ACCESS project type, a per-account QoS proxy is inferred from annualised usage and mapped to the four ACCESS-style tiers (Explore, Discover, Accelerate, Maximize), giving the simulator realistic account-scale heterogeneity (Appendix~\ref{app:qos}). Descriptively, the Anvil workload is bursty in time and highly skewed across accounts (Figure~\ref{fig:datastats}): demand concentrates in working hours, and a small number of accounts submit a disproportionate share of jobs --- both features that any realistic synthetic workload must preserve.

\begin{figure}[htbp]\centering
\begin{subfigure}{0.46\linewidth}\centering
\includegraphics[width=\linewidth]{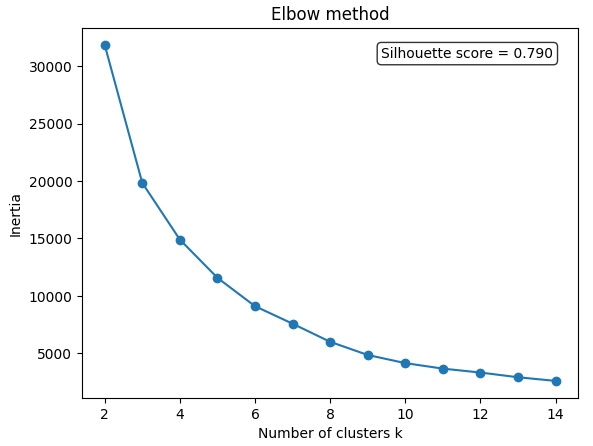}
\caption{Elbow/inertia}
\end{subfigure}\hfill
\begin{subfigure}{0.46\linewidth}\centering
\includegraphics[width=\linewidth]{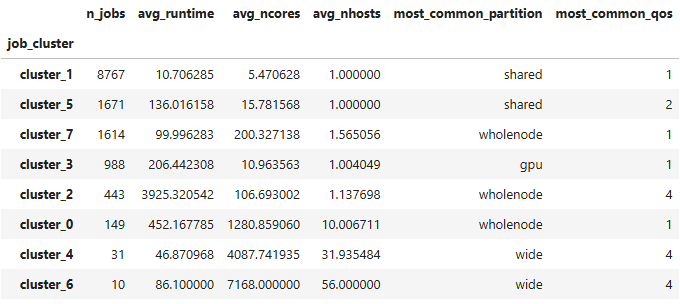}
\caption{Cluster statistics}
\end{subfigure}
\caption{Clustering diagnostics used to choose and validate the number of workload groups before synthetic generation.}
\label{fig:clustering}
\end{figure}

\begin{figure}[htbp]\centering
\includegraphics[width=0.74\linewidth]{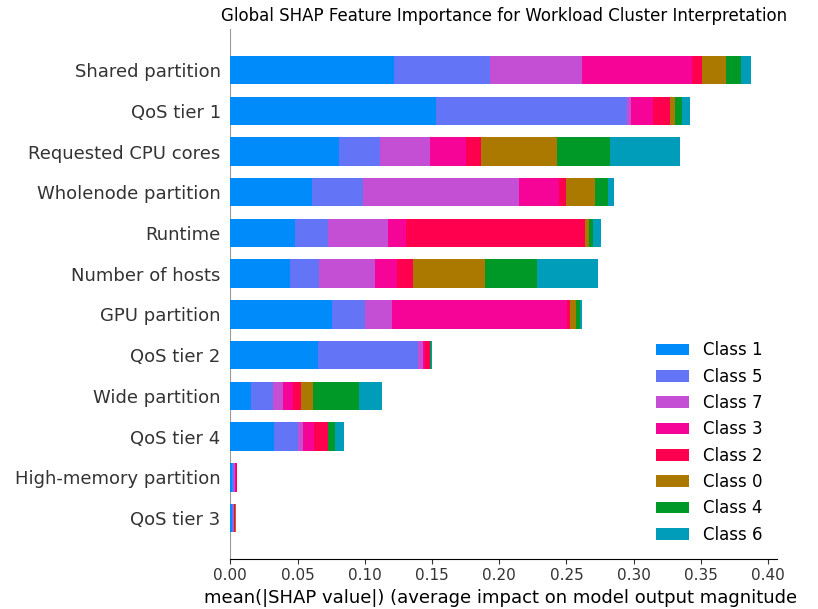}
\caption{SHAP feature importance for the surrogate workload classifier. The learned clusters are driven by scheduling-relevant features (partition, cores, runtime, hosts), confirming the synthetic traces preserve realistic structure.}
\label{fig:shap}
\end{figure}

\paragraph{Synthetic workload construction.} The selected Anvil months are low-contention relative to Anvil-scale capacity, so replaying them directly is useful for realism but cannot stress a scheduler. We therefore learn the workload's structure and regenerate controlled synthetic traces. Jobs are first grouped by clustering on scheduling-relevant features (requested cores, hosts, runtime, partition, QoS), with cluster quality checked by the elbow/inertia and silhouette diagnostics (Figure~\ref{fig:clustering}); a surrogate classifier is then trained to reproduce the cluster labels, and its SHAP attributions confirm the groups are driven by genuine scheduling features --- the shared and whole-node partitions, requested cores, runtime, and host count --- rather than artefacts (Figure~\ref{fig:shap})~\citep{shap_tools}. Synthetic jobs are sampled from this learned structure and given arrival times under three regimes: an \emph{even-fill} regime (denser, regular demand), a \emph{bursty-real} regime (preserving the real burst structure), and a \emph{bursty-fill} regime that adds jobs into quiet periods for stress tests, with volume multipliers to tune pressure. The framework supports all three, but no results using the even-fill regime are reported here: the contention experiments below use the bursty-real and bursty-fill traces. The synthetic traces thus preserve realistic job structure while letting us control the one thing the real months lack --- contention.

The framework replays the same workload under four rule families --- FIFO (transparent baseline), Slurm-like weighted priority (operational heuristic), multi-objective-tuned weights, and an offline ILP benchmark --- and, in Experiment~4, under federation policies. Outcomes are scored on system efficiency (utilisation, throughput, makespan), user-facing delay (waiting time, turnaround, slowdown), and fairness (account/job-size slowdown gap and Jain indices). Requested walltime is treated as both a scheduling input and a behavioural signal, isolated by replaying under actual-runtime versus requested-walltime occupation modes. We address four research questions: how weights map to objectives (RQ1); how far heuristics sit from the offline benchmark (RQ2); how strategic reporting creates externalities (RQ3); and how federation balances global gains against local-user protection (RQ4).

\paragraph{Scheduling rules and occupation modes.} The rule families compared span the full spectrum from transparent to optimal (Table~\ref{tab:rules}). Because requested walltime is simultaneously a scheduling input and a behavioural signal, we replay each workload under three occupation modes to separate the two effects: (i) \emph{actual runtime} used for both priority and resource holding (the best-information case); (ii) \emph{requested walltime for priority, actual runtime for occupation} (isolating the ordering distortion); and (iii) \emph{requested walltime for both} (the pessimistic blocking case that also captures over-reservation). Comparing modes lets us attribute outcomes to queue re-ordering versus resource over-holding.

\begin{table}[htbp]\centering
\caption{Scheduling rules compared across the simulation study.}
\label{tab:rules}
\small
\begin{tabularx}{\linewidth}{@{}lX@{}}
\toprule
\textbf{Rule} & \textbf{Role} \\
\midrule
FIFO & Transparent baseline; easy to explain, optimises nothing. \\
Slurm-like priority & Operationally realistic weighted score (age, fairshare, size, QoS). \\
Tuned priority (pymoo/NSGA-II) & Weights searched on the efficiency--fairness trade-off surface. \\
Offline ILP (PyJobShop) & Full-information welfare benchmark (weighted flow time / value). \\
Federation broker & Cross-system placement policies (Experiment 4). \\
\bottomrule
\end{tabularx}
\end{table}

\paragraph{Evaluation metrics.} Outcomes are scored on three axes. \emph{System efficiency}: platform utilisation, throughput, and makespan. \emph{User-facing delay}: waiting time (submission to start), turnaround (submission to completion), and slowdown (turnaround divided by runtime, which normalises delay by job length). \emph{Fairness}: the account-level and job-size slowdown gap (max$-$min), and Jain's index applied both to allocated CPU service (a desirable quantity) and to delay (a cost). A good schedule combines low average slowdown with a high slowdown-Jain index; a high Jain index alone can simply mean everyone is served equally badly, which is why we always report fairness alongside the average level of delay. Formal definitions of every metric are collected in Appendix~\ref{app:metrics}.

\subsection{Experiment 1 --- Priority weights as policy levers}
Treating the Slurm weights over age, fairshare, job size, and QoS as decision variables, we search the trade-off surface with the NSGA-II genetic algorithm via pymoo~\citep{blank2020pymoo,pymoo2026,nsga2002}, scoring each candidate on utilisation, waiting time, slowdown, and account-level slowdown gap. On the low-contention real trace, weights mainly redistribute delay; under synthetic contention they become genuine policy levers. The search returns a Pareto set of non-dominated policies (Table~\ref{tab:exp1}): no single configuration is best on every objective --- one maximises utilisation and fairness, another minimises waiting time, another slowdown, with a balanced compromise in between. The operational implication is that choosing weights \emph{is} setting policy, and there is no universally correct setting.

\begin{table}[htbp]\centering
\caption{Representative Pareto-optimal weight policies on the synthetic filled workload (Experiment 1). No policy dominates on all objectives.}
\label{tab:exp1}
\small
\begin{tabularx}{\linewidth}{@{}lYYYY@{}}
\toprule
\textbf{Policy} & \textbf{Utilisation} & \textbf{Avg wait (min)} & \textbf{Avg slowdown} & \textbf{Account gap} \\
\midrule
P1 (utilisation / fairness) & 0.874 & 379.5 & 13.32 & 552.1 \\
P5 (waiting time) & 0.871 & 342.2 & 8.57 & 565.0 \\
P6 (balanced) & 0.869 & 358.0 & 8.22 & 561.5 \\
P8 (slowdown) & 0.865 & 354.8 & 8.19 & 561.6 \\
\bottomrule
\end{tabularx}
\end{table}

To expose \emph{why} multi-objective tuning matters, we also ran four single-objective searches, each targeting only one metric (Table~\ref{tab:exp1-single}). A reading note: because the evolutionary search evaluates a finite set of candidate weight configurations, each row reports the best configuration \emph{found} for its target, not a guaranteed optimum of that metric --- which is why the slowdown-focused run happened to land on a configuration with slightly higher utilisation (0.919) than the utilisation-focused run (0.914). The qualitative pattern is nonetheless clear: the utilisation-focused configuration incurs a visible cost in waiting time, slowdown, and account fairness, while the delay- and fairness-oriented configurations sacrifice very little utilisation while improving user-facing outcomes. In the multi-objective run, extreme policies of this kind do not survive as Pareto solutions because waiting time and slowdown are protected simultaneously. The practical lesson is direct: optimising one scheduling goal in isolation quietly damages the others, so operators need explicit multi-objective guidance rather than a single default weight vector.

\begin{table}[htbp]\centering
\caption{Single-objective priority-weight tuning on the synthetic filled workload (Experiment 1). Each row reports the best configuration the search found for one target metric; note the collateral damage to the others.}
\label{tab:exp1-single}
\small
\begin{tabularx}{\linewidth}{@{}lYYYY@{}}
\toprule
\textbf{Objective optimised} & \textbf{Utilisation} & \textbf{Avg wait (min)} & \textbf{Avg slowdown} & \textbf{Account gap} \\
\midrule
Maximise utilisation & 0.914 & 1058.5 & 70.00 & 833.0 \\
Minimise waiting time & 0.909 & 965.6 & 62.17 & 768.2 \\
Minimise slowdown & 0.919 & 961.0 & 62.13 & 765.8 \\
Minimise account gap & 0.911 & 960.4 & 61.83 & 764.2 \\
\bottomrule
\end{tabularx}
\end{table}

\subsection{Experiment 2 --- Practical rules versus the offline benchmark}
We replay a high-pressure workload under FIFO, Slurm-like size-priority, a tuned policy, and an offline ILP (minimising weighted flow time) implemented with PyJobShop~\citep{lan2025pyjobshop,pyjobshopdocs2026}. FIFO is clearly worst. The tuned and size-priority heuristics actually match or slightly beat the offline ILP on average waiting time --- unsurprising, since the ILP optimises weighted flow time rather than raw wait, and is itself solved only to a small optimality gap (Section~\ref{sec:model-benchmark}) --- while the ILP retains the lowest average slowdown; the heuristics improve sharply on FIFO in delay and fairness --- at a slight cost in raw utilisation --- without dominating the full-information benchmark overall (Table~\ref{tab:exp2}). The message for operators is that transparent, Slurm-compatible heuristics are ``good enough'' once tuned: the near-term priority is to tune existing schedulers, not to replace them.

\begin{table}[htbp]\centering
\caption{Practical rules versus the offline ILP benchmark on the synthetic filled workload (Experiment 2).}
\label{tab:exp2}
\small
\begin{tabularx}{\linewidth}{@{}lYYYY@{}}
\toprule
\textbf{Rule} & \textbf{Utilisation} & \textbf{Avg wait (min)} & \textbf{Avg slowdown} & \textbf{Account gap} \\
\midrule
Offline ILP (best found) & 0.766 & 83.5 & 1.76 & 17.4 \\
FIFO & 0.755 & 112.7 & 4.11 & 102.1 \\
Size-priority & 0.737 & 81.1 & 1.86 & 17.4 \\
Tuned (P5) & 0.737 & 82.8 & 2.05 & 24.1 \\
\bottomrule
\end{tabularx}
\end{table}

\begin{figure}[htbp]\centering
\begin{subfigure}{0.48\linewidth}\centering
\includegraphics[width=\linewidth]{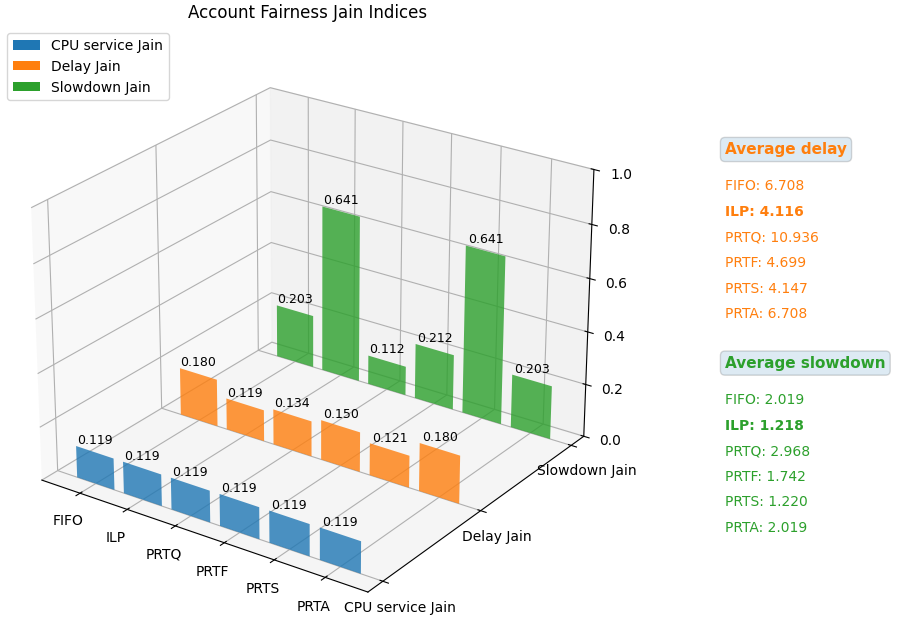}
\caption{Real trace (low contention)}
\end{subfigure}\hfill
\begin{subfigure}{0.48\linewidth}\centering
\includegraphics[width=\linewidth]{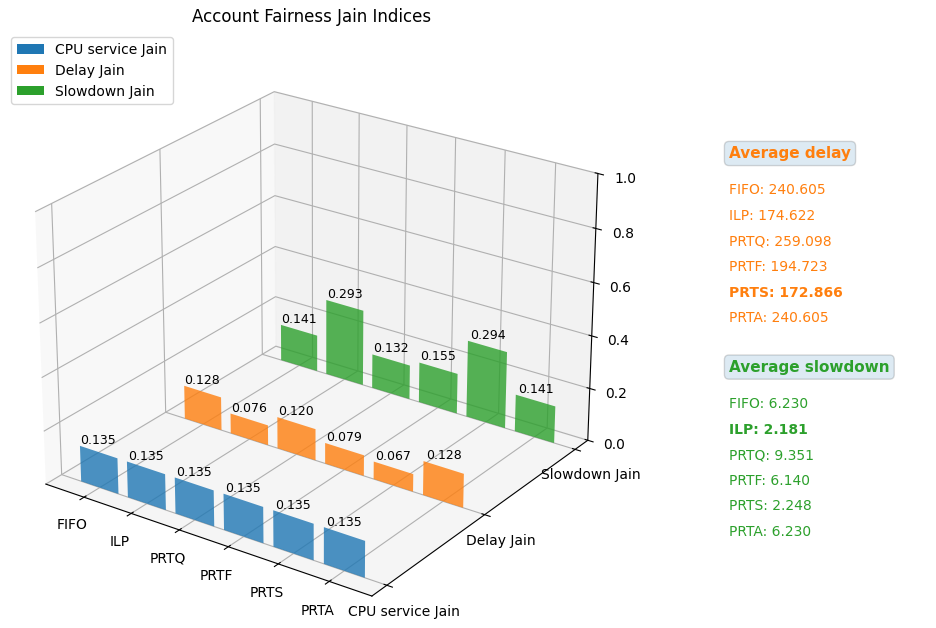}
\caption{Synthetic filled (high contention)}
\end{subfigure}
\caption{Account-level fairness (Jain index) across scheduling rules. Under low contention (left) the rules are nearly indistinguishable; under contention (right) the distributional consequences of policy choice emerge.}
\label{fig:jain}
\end{figure}

It is equally instructive to run the same comparison on the \emph{real} Anvil trace, which is low-contention: there, platform utilisation is pinned at roughly 0.25 for \emph{every} rule, and ILP, FIFO and the priority variants become nearly indistinguishable on delay and fairness (Table~\ref{tab:exp2-original}). When there are rarely enough queued jobs to compete, the scheduler has little to decide, so priority weights mostly redistribute small amounts of delay rather than changing aggregate outcomes. This is not a null result but a \emph{scoping} result: it demonstrates that scheduling policy only bites under contention, which is exactly why the controlled synthetic stress tests --- not the sparse real months --- are the right instrument for evaluating allocation mechanisms, and why a high-contention \emph{real} UK trace (FAIRC-1) would be so valuable. The fairness picture (Figure~\ref{fig:jain}) tells the same story: under load the distributional differences between rules become visible, with FIFO and a QoS-heavy rule producing the largest account-level imbalance and size-priority the most even service.

\begin{table}[htbp]\centering
\caption{Practical rules on the low-contention \emph{real} Anvil trace (Experiment 2). Utilisation is capped by available demand ($\approx$0.25), so the rules barely differ --- scheduling policy only bites under contention.}
\label{tab:exp2-original}
\small
\begin{tabularx}{\linewidth}{@{}lYYYY@{}}
\toprule
\textbf{Rule} & \textbf{Utilisation} & \textbf{Avg wait (min)} & \textbf{Avg slowdown} & \textbf{Account gap} \\
\midrule
Offline ILP (best found) & 0.252 & 2.26 & 1.11 & 6.8 \\
FIFO & 0.252 & 4.41 & 1.53 & 30.3 \\
Size-priority & 0.252 & 2.27 & 1.11 & 6.8 \\
Fairshare-priority & 0.252 & 3.14 & 1.36 & 25.6 \\
QoS-priority & 0.252 & 5.32 & 1.78 & 50.1 \\
\bottomrule
\end{tabularx}
\end{table}

\subsection{Experiment 3 --- Strategic reporting and scheduling externalities}
\begin{figure}[htbp]\centering
\includegraphics[width=0.72\linewidth]{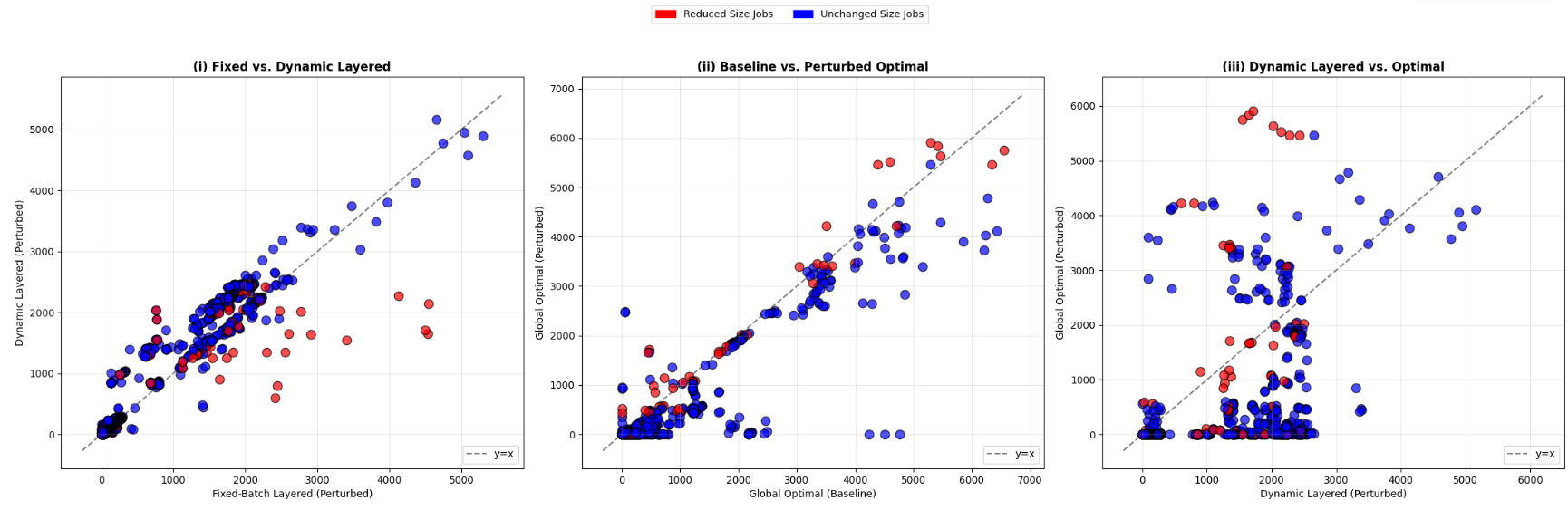}
\caption{Job-level waiting times under runtime perturbation. Red points are jobs whose runtimes were reduced; blue points are unchanged. Blue points moving off the diagonal show that unchanged jobs are delayed or advanced by queue re-ranking --- a system-wide externality.}
\label{fig:walltime}
\end{figure}

Two sub-experiments probe private information. In the \textbf{runtime-perturbation} study, reducing the runtimes of 20\% of jobs improves aggregate delay (weighted flow time falls from 3.85M to 3.42M; average wait from 222 to 195 minutes), but Figure~\ref{fig:walltime} shows the improvement is not uniform: many \emph{unchanged} jobs (blue) also shift because shorter occupation releases capacity and the scheduler re-ranks the queue. Runtime information thus creates a \textbf{queue externality} --- one user's report changes other users' waiting times. The aggregate effect is summarised in Table~\ref{tab:exp3-runtime}: the fixed-batch schedule, which keeps the original job order but uses the shorter runtimes, improves \emph{more} than the dynamic scheduler that re-ranks the queue, showing that some of the capacity released by shorter jobs is dissipated by the reordering itself, and the offline benchmark reveals the remaining gap to full-information scheduling.

\begin{table}[htbp]\centering
\caption{Aggregate effect of reducing the runtimes of 20\% of jobs (Experiment 3). Lower is better; ``global'' rows are the offline full-information benchmark.}
\label{tab:exp3-runtime}
\small
\begin{tabularx}{\linewidth}{@{}lYY@{}}
\toprule
\textbf{Schedule} & \textbf{Weighted flow time} & \textbf{Avg wait (min)} \\
\midrule
Baseline, dynamic layered & 3{,}853{,}622 & 222.0 \\
Baseline, offline global & 2{,}926{,}142 & 89.5 \\
Perturbed, dynamic layered & 3{,}416{,}793 & 194.6 \\
Perturbed, fixed-batch & 3{,}264{,}625 & 172.8 \\
Perturbed, offline global & 2{,}513{,}687 & 65.6 \\
\bottomrule
\end{tabularx}
\end{table}

\begin{figure}[htbp]\centering
\includegraphics[width=0.66\linewidth]{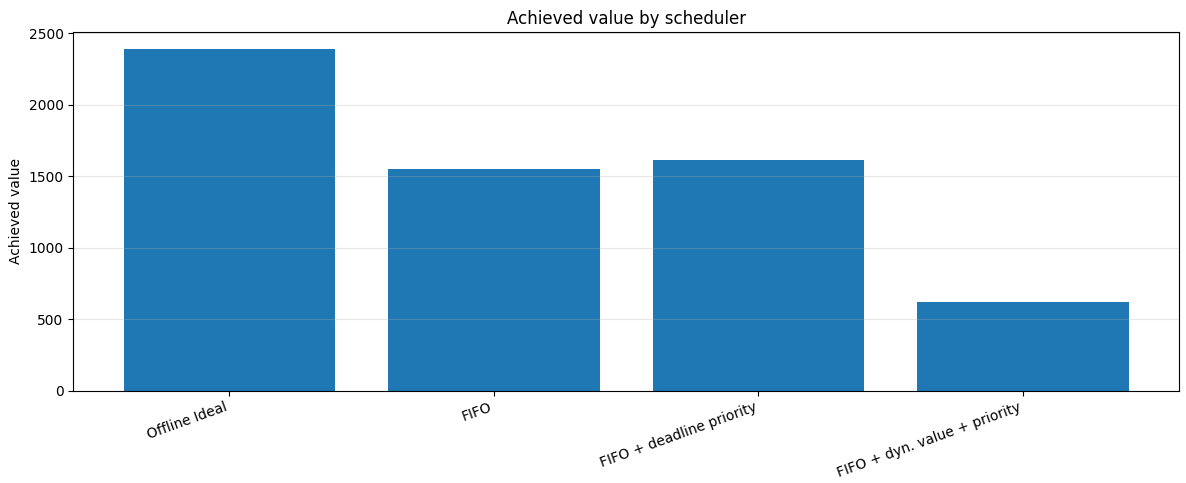}
\caption{Achieved scientific value by scheduler. The offline benchmark preserves the full available value; value-blind online rules recover only around two-thirds.}
\label{fig:value}
\end{figure}

In the \textbf{deadline-and-value} study, each job carries a soft deadline and a scientific value; the offline benchmark preserves all 2388 value units and completes all 500 jobs on time, whereas the online rules recover only 1552--1612 units and finish around 340 jobs on time --- despite having \emph{lower} average waiting time and higher utilisation (Table~\ref{tab:exp3value}, Figure~\ref{fig:value}). Operational metrics can therefore be misleading: a schedule may look efficient while destroying a large fraction of scientific value. This motivates value-aware prioritisation --- with the caveat that if value or urgency is self-reported, the mechanism must be incentive-compatible.

\begin{table}[htbp]\centering
\caption{Deadline-value scheduling under the full-deadline value metric (Experiment 3). Online rules look efficient on wait/utilisation yet lose substantial value.}
\label{tab:exp3value}
\small
\begin{tabularx}{\linewidth}{@{}lYYYY@{}}
\toprule
\textbf{Scheduler} & \textbf{On-time jobs} & \textbf{Avg wait (min)} & \textbf{Avg slowdown} & \textbf{Utilisation} \\
\midrule
Offline PyJobShop (best found) & 500 & 1055.3 & 1.61 & 0.614 \\
FIFO & 341 & 557.4 & 47.04 & 0.824 \\
FIFO + deadline priority & 340 & 563.3 & 46.72 & 0.819 \\
FIFO + dynamic value priority & 339 & 562.4 & 46.33 & 0.820 \\
\bottomrule
\end{tabularx}
\end{table}

\subsection{Experiment 4 --- Federation as routing under congestion}
Two autonomous systems $A$ and $B$ each run a local FIFO scheduler; a broker decides at submission whether a job stays home or moves, under policies of increasing coordination: no federation, naive (go to the lowest estimated completion time), switching-cost (a delay ``toll'' on remote execution), minimum-gain (move only if the saving exceeds a threshold), and utilisation-protected (move only if the receiver has spare capacity). The natural lens is a congestion-game \emph{analogy}: naive placement resembles unpriced selfish routing, in which adding capacity to an uncoordinated network can worsen outcomes (Braess's paradox). By analogy, uncoordinated placement may concentrate load on receiving systems and create risks for their local users. Appendix~\ref{app:federation} gives the full policy definitions and the complete results across all four scenarios; the main text reports selected rows.

Federation helps most under asymmetric load (Table~\ref{tab:fed}): naive routing cuts average wait from 127.8 to 4.0 minutes. But naive is best on aggregate delay precisely because it is selfish --- it consumes shared spare capacity aggressively and concentrates load on the receiving system. The switching-cost, minimum-gain, and utilisation-protected rules trade a little global delay to filter weak transfers and steer movement into the genuine relief direction ($B\!\to\!A$), protecting the receiving system's local users. For example, utilisation protection raises average wait from 4.0 to 9.0 minutes but cuts $A\!\to\!B$ transfers from 5{,}974 to 1{,}748 while \emph{improving} slowdown from 1.39 to 1.18. The finding is deliberately honest: on aggregate metrics naive currently looks best. The federation results are \emph{consistent with} the selfish-routing analogy --- uncoordinated placement reduces aggregate delay while generating large, sometimes bidirectional, transfer volumes that may concentrate load on receiving systems --- but the present simulations do not directly quantify harm to local users; they provide directional evidence for the need to examine congestion and local-service risks. The case for coordination therefore rests on protecting local service and stability as the federation scales, and quantifying local harm precisely is part of the remaining work.

\begin{table}[htbp]\centering
\caption{Asymmetric-load federation (system $A$ idle, $B$ busy). Naive minimises delay but moves work indiscriminately; coordinated rules protect the receiver.}
\label{tab:fed}
\small
\begin{tabularx}{\linewidth}{@{}lYYYY@{}}
\toprule
\textbf{Policy} & \textbf{Avg wait (min)} & \textbf{Avg slowdown} & \textbf{$A\!\to\!B$} & \textbf{$B\!\to\!A$} \\
\midrule
No federation & 127.8 & 13.33 & 0 & 0 \\
Naive & 4.0 & 1.39 & 5{,}974 & 13{,}402 \\
Switching cost & 18.9 & 2.17 & 739 & 22{,}466 \\
Utilisation-protected & 9.0 & 1.18 & 1{,}748 & 8{,}751 \\
\bottomrule
\end{tabularx}
\end{table}

Federation still helps under \emph{balanced} high load, because two systems are rarely congested in lockstep (Table~\ref{tab:fed-balanced}); one temporarily has usable capacity while the other spikes. Here naive again gives the lowest average wait, but its unpriced character is clearer: it moves large volumes in \emph{both} directions ($A\!\to\!B$ and $B\!\to\!A$), whereas switching cost filters weak transfers and lets the direction of net relief emerge. In the heterogeneous case, where system~$B$ carries the GPUs, naive over-uses the specialised resource; the toll relieves that pressure and lets CPU-suitable work flow back to~$A$.

\begin{table}[htbp]\centering
\caption{Balanced high-load federation: both systems under comparable pressure. Naive minimises wait but moves work in both directions; switching cost concentrates movement.}
\label{tab:fed-balanced}
\small
\begin{tabularx}{\linewidth}{@{}llYYYY@{}}
\toprule
\textbf{Scenario} & \textbf{Policy} & \textbf{Avg wait} & \textbf{Avg slow.} & \textbf{$A\!\to\!B$} & \textbf{$B\!\to\!A$} \\
\midrule
\multirow{3}{*}{Homogeneous} & No federation & 133.8 & 16.50 & 0 & 0 \\
 & Naive & 51.3 & 8.88 & 11{,}770 & 13{,}497 \\
 & Switching cost & 71.6 & 11.81 & 7{,}179 & 14{,}692 \\
\addlinespace
\multirow{3}{*}{Heterogeneous} & No federation & 20.0 & 3.83 & 0 & 0 \\
 & Naive & 1.3 & 1.21 & 14{,}502 & 9{,}062 \\
 & Switching cost & 9.0 & 1.28 & 7{,}633 & 12{,}073 \\
\bottomrule
\end{tabularx}
\end{table}

\begin{figure}[htbp]\centering
\includegraphics[width=0.72\linewidth]{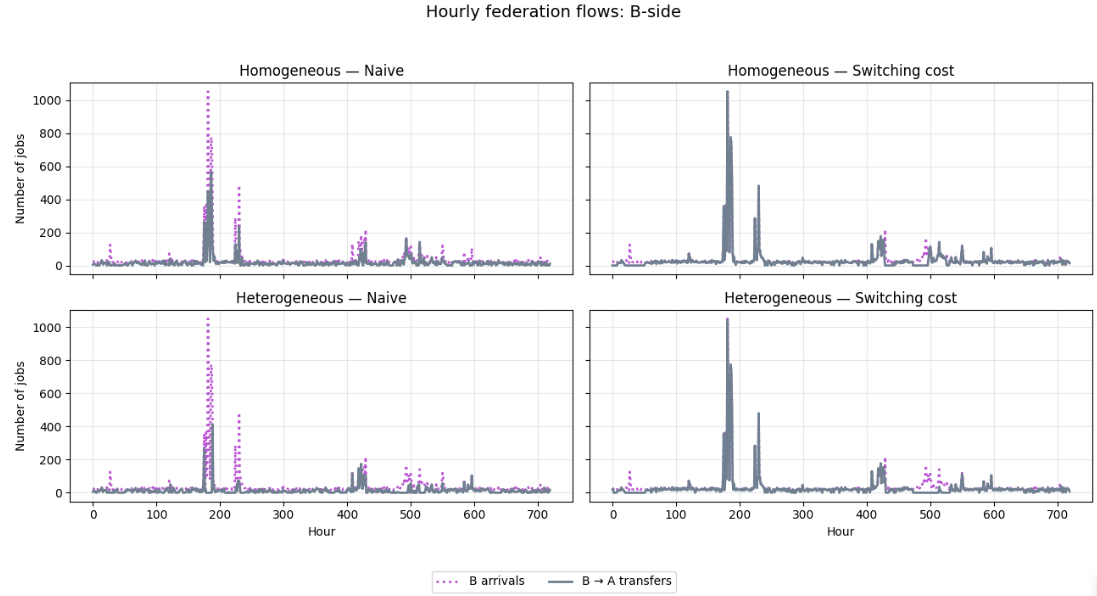}
\caption{Cross-system transfers over time under naive versus switching-cost routing (asymmetric load). The switching cost suppresses weak $A\!\to\!B$ movement and concentrates $B\!\to\!A$ relief around system $B$'s demand spike --- behavioural evidence of route regulation rather than mere volume reduction.}
\label{fig:fed}
\end{figure}

Finally, the minimum-gain and utilisation-protected families (Table~\ref{tab:fed-conservative}) show that protection is a \emph{dial}, not a binary choice between isolation and a free-for-all. Minimum-gain removes low-benefit transfers, keeping most of the delay benefit while sharply cutting cross-system traffic; adding a switching cost makes movement more \emph{directional}, concentrating it in the genuine relief direction rather than simply reducing it. The utilisation-protected policy is designed to shield a receiving system's own users, limiting transfers when the receiver is under pressure at the cost of a little more delay; the relative effect of the two families varies across scenarios (minimum-gain yields fewer transfers in some cases, as Table~\ref{tab:fed-conservative} shows), so differences in transfer volumes should be read as directional rather than as a fixed ordering. Figure~\ref{fig:fed} visualises the effect --- switching cost suppresses weak $A\!\to\!B$ transfers while allowing strong $B\!\to\!A$ relief around $B$'s demand spike.

\begin{table}[htbp]\centering
\caption{Conservative federation rules (heterogeneous scenarios). Coordination trades a little delay for markedly less --- and more directional --- cross-system movement, protecting receiving systems.}
\label{tab:fed-conservative}
\small
\begin{tabularx}{\linewidth}{@{}llYYYY@{}}
\toprule
\textbf{Scenario} & \textbf{Policy} & \textbf{Avg wait} & \textbf{Avg slow.} & \textbf{$A\!\to\!B$} & \textbf{$B\!\to\!A$} \\
\midrule
\multirow{3}{*}{Low $A$ / high $B$} & Naive & 0.2 & 1.00 & 6{,}822 & 9{,}108 \\
 & Minimum-gain & 0.3 & 1.00 & 2{,}801 & 7{,}080 \\
 & Utilisation-protected & 0.5 & 1.00 & 3{,}051 & 7{,}682 \\
\addlinespace
\multirow{3}{*}{Balanced} & Naive & 1.3 & 1.21 & 14{,}502 & 9{,}062 \\
 & Minimum-gain & 1.0 & 1.16 & 7{,}514 & 5{,}125 \\
 & Utilisation-protected & 3.8 & 1.50 & 7{,}454 & 5{,}075 \\
\bottomrule
\end{tabularx}
\end{table}

\subsection{Summary of findings}
The simulations support four roadmap-relevant conclusions: (i) scheduler weights are policy levers with genuine trade-offs, so operators need guidance, not a single default (\textbf{FAIRC-4}); (ii) tuned transparent heuristics approach the offline benchmark, so tune before replacing (\textbf{FAIRC-4}); (iii) private runtime and value information create system-wide externalities that only proper measurement can expose (\textbf{FAIRC-8}, \textbf{FAIRC-1/2}), and, under a synthetic value model, value-blind scheduling can leave a large fraction of scientific value unrealised (\textbf{FAIRC-3}); and (iv) federation is valuable but must be coordinated to avoid congestion-style harm (\textbf{FAIRC-5}), which in turn needs fair cross-system exchange rates (\textbf{FAIRC-6}). All quantitative results on contention derive from synthetic stress tests (a clearly-labelled sandbox), because a high-contention UK trace was not available --- the strongest single argument for FAIRC-1.

\paragraph{Configuration and gap analysis.} The federation study abstracts to two autonomous systems (each 300 nodes $\times$ 128 CPU cores, with one GPU per node on the heterogeneous system~$B$; 30-day synthetic traces), using a switching-cost penalty as a stand-in for network transfer cost and not modelling inter-site data movement or authentication. Experiment 1 and the runtime-perturbation study use a representative 7{,}000-job high-pressure window (simulated platform: 300 nodes $\times$ 128 cores, i.e.\ 38{,}400 CPU cores), while the deadline-value study uses a 500-job window to keep the offline optimisation tractable. All reported simulation figures are point estimates from these representative windows rather than averages over multiple random seeds, and should be read as indicative magnitudes rather than precise expected values. The principal gap is the absence of a high-contention \emph{real} UK trace: the contention results are therefore counterfactual sandbox experiments, and validating them on pseudonymised UK job logs --- enabled by FAIRC-1 --- is the necessary next step before any operational adoption. Implementation and reproducibility notes for the whole framework are collected in Appendix~\ref{app:software}.

% =================== RECOMMENDATIONS ===================
\section{Recommendations for the NFCS Roadmap}
\label{sec:recs}

The eight recommendations below follow the NFCS output guidance: each carries the FAIRC short code, a Must/Should/Could-Have category, recommendation text with rationale, an indicative timescale within the five-year roadmap, and a resource estimate. \emph{FTE and cost figures are first-pass planning estimates for discussion, drawing on operational benchmarks gathered in interviews, and should be confirmed before adoption.} They are ordered by suggested implementation sequence.

\begin{recbox}{FAIRC-1 --- Unified data-collection standard}{Must Have}
\textbf{Recommendation.} Establish a common, mandatory data-collection and reporting standard for all federated systems, capturing per job a minimal agreed schema: requested versus consumed resources (cores, memory, GPU, walltime), queue/partition and QoS, account and Research-Council attribution, completion status, and --- where a request is refused --- an anonymised reason-for-rejection. Reporting should be automated, pseudonymised at source, and aggregated through the planned single-pane-of-glass access layer.\\[4pt]
\textbf{Rationale.} Every other recommendation, and any future allocation mechanism, depends on data that does not currently exist in a consistent form. Across our interviews the same gap recurred: operators ``don't measure enough or anything now,'' outputs are weakly tracked, and monitoring quality varies widely. Our own simulation work was forced onto public US data precisely because UK job logs could not be obtained on workable terms. Without a shared schema the federation cannot answer whether a resource is used for the purpose it was granted, and productively. This is the foundational, lowest-regret investment, and underpins interoperability across heterogeneous National Compute Resources.\\[4pt]
\textbf{Timescale.} Year 1, 12 months. Foundational --- a dependency for FAIRC-2, 3, 5 and 8. Pilot the schema on 2--3 willing systems in the first 6 months, then extend.\\[2pt]
\textbf{Resources.} $\sim$1.5 FTE (1.0 Research Software Engineer; 0.5 data/policy analyst); $\sim$\pounds140,000. Aligns with the platform-services / accounting layer of the disaggregated stack.
\end{recbox}

\begin{recbox}{FAIRC-2 --- Occupancy-vs-utilisation KPI}{Must Have}
\textbf{Recommendation.} Adopt a standard efficiency metric across federated systems, defined as realised utilisation divided by reserved occupancy (e.g.\ delivered FLOP or core-hours against reserved core-hours), reported alongside existing occupancy-based accounting and paired with at least one outcome indicator so that ``productive use'' is measured rather than assumed.\\[4pt]
\textbf{Rationale.} Current accounting charges for occupancy, not utilisation; a system can show high occupancy and low utilisation (reserved-but-idle), which is pure opportunity cost. DSIT and UKRI repeatedly framed their open question as defining and measuring ``productive use'' and ROI; DiRAC already treats 85--90\% utilisation as an economic necessity. A standard occupancy--utilisation signal gives the federation a comparable basis for value-for-money judgements and exposes the over-reservation that walltime padding produces. JASMIN shows storage occupancy can itself be the binding constraint, so metrics should extend to storage efficiency where appropriate.\\[4pt]
\textbf{Timescale.} Year 1--2, 12 months. Depends on FAIRC-1 telemetry. Storage-bound systems should report a parallel storage-occupancy measure.\\[2pt]
\textbf{Resources.} $\sim$0.5 FTE (data analyst) plus operator effort to expose telemetry; $\sim$\pounds50,000. Builds on existing monitoring rather than new infrastructure.
\end{recbox}

\begin{recbox}{FAIRC-4 --- Tune Slurm weights as policy levers}{Should Have}
\textbf{Recommendation.} Treat scheduler priority weights (age, fairshare, job size, QoS) as policy instruments, and publish practical guidance mapping common policy objectives --- utilisation, responsiveness, slowdown, account fairness --- to weight configurations. Retain existing schedulers; do not mandate replacement.\\[4pt]
\textbf{Rationale.} Our simulations show that, under contention, priority-weight choices act as genuine allocation policy: no single configuration is best on all objectives --- the tuning yields a set of trade-off options, each best on a different goal (utilisation, waiting time, fairness), with no single overall winner. Tuned, transparent Slurm-compatible heuristics land close to the offline full-information benchmark, so they are ``good enough'' for most current loads and should be tuned, not replaced --- a low-cost, low-risk improvement operators can adopt immediately.\\[4pt]
\textbf{Timescale.} Year 1, 9 months. Short-term; benefits from FAIRC-1 for validation but does not require it.\\[2pt]
\textbf{Resources.} $\sim$0.5 FTE (postdoctoral researcher) with operator review; $\sim$\pounds45,000. Directly usable by Tier-2/Tier-3 operators and CCEs.
\end{recbox}

\begin{recbox}{FAIRC-7 --- Tiered rapid access and CCEs}{Should Have}
\textbf{Recommendation.} Reduce the administrative barrier on two complementary fronts: introduce a lightweight rapid-access route (``Seedcorn / Bright Idea'') for small, exploratory, or early-career requests alongside standard peer review; and resource Community Centres of Excellence (CCEs) as the community-facing coordination layer, brokering allocations across systems and providing training and RSE support.\\[4pt]
\textbf{Rationale.} Manual peer review is the binding administrative constraint, not hardware --- operators described allocation reviews consuming 60--100 person-hours and DiRAC assessing $\sim$67 applications per round under a ``double jeopardy'' process. Uniform heavyweight review does not scale and disproportionately excludes early-career researchers and non-traditional disciplines. A rapid tier lowers the transactional barrier (monitored via FAIRC-1 rather than gated by review); CCEs address the structural barrier by pairing brokerage with training and RSE support, complementing the automated mechanisms in FAIRC-1, 2 and 8 with the human coordination layer federation needs.\\[4pt]
\textbf{Timescale.} Phased Years 1--3. Rapid route: Year 1, $\sim$6 months. CCE coordination: Years 2--3, alongside the NCR rollout.\\[2pt]
\textbf{Resources.} Rapid route $\sim$0.3 FTE, $\sim$\pounds30,000/year; CCE coordination leverages UKRI's planned CCE investment (define the allocation-brokerage interface, $\sim$0.3 FTE, $\sim$\pounds30,000).
\end{recbox}

\begin{recbox}{FAIRC-6 --- Cross-system benchmarking \& exchange rates}{Should Have}
\textbf{Recommendation.} Adopt a federation-wide benchmarking and exchange-rate framework so that allocations are portable across heterogeneous systems, using domain-representative workloads rather than a single synthetic benchmark, with published conversion factors.\\[4pt]
\textbf{Rationale.} Portable allocation --- central to federation and to the ACCESS-style credit model --- requires a defensible way to compare a core-hour on one system with a core-hour on another. The WLCG's move from the HS06 benchmark to HEPScore --- which uses containerised real experiment workloads and is reproducible to better than 1\% across x86 and ARM processors --- is the proven template~\citep{giordano2023hepscore}; ACCESS already operationalises performance-equivalence multipliers. Without a transparent exchange basis, cross-system mobility stalls or embeds hidden inequities between communities and architectures (notably CPU versus GPU).\\[4pt]
\textbf{Timescale.} Year 1--2, 12 months. Medium-term; a precondition for portable credits and for FAIRC-5. Benefits from FAIRC-1.\\[2pt]
\textbf{Resources.} $\sim$1.0 FTE (benchmarking engineer) with domain-community input; $\sim$\pounds95,000. Aligns with WLCG/HEPScore, ACCESS, and EuroHPC/Waldur accounting.
\end{recbox}

\begin{recbox}{FAIRC-8 --- Ex-post verification \& usage history}{Should Have}
\textbf{Recommendation.} Introduce ex-post verification comparing requested against consumed resources, feeding a transparent account-level usage-history state into future allocation and priority, so that persistent over-requesting or sustained under-use is gently penalised and accurate forecasting rewarded. Keep penalties graduated and appealable.\\[4pt]
\textbf{Rationale.} This codifies practice operators already use informally: IRIS-SCRC accounts for prior under/over-use in future allocations, and DiRAC triggers follow-up review below 50\% (small) or 80\% (large) utilisation and decays credits quarterly. Our model represents this as a usage-history state that links a user's past behaviour to their future priority. Formalising it federation-wide discourages walltime padding and hoarding --- the over-reservation FAIRC-2 measures --- while a sandbox or experience-based grace period protects newcomers who genuinely cannot estimate their needs. Verification depends on FAIRC-1 telemetry.\\[4pt]
\textbf{Timescale.} Year 2, 9 months. Medium-term. Depends on FAIRC-1; complements FAIRC-2.\\[2pt]
\textbf{Resources.} $\sim$1.0 FTE (0.5 mechanism-design researcher; 0.5 RSE); $\sim$\pounds90,000. Extends IRIS-SCRC and DiRAC practice.
\end{recbox}

\begin{recbox}{FAIRC-3 --- Pilot value-aware prioritisation (PVR)}{Could Have}
\textbf{Recommendation.} Pilot an allocation mechanism that accounts for the scientific value lost when time-sensitive jobs wait, using a Projected Value Remaining (PVR) measure that models the non-increasing value of a job over time, evaluated against current schedulers before any production use. Design the interface so users can express time-sensitivity simply, without disclosing complex private information.\\[4pt]
\textbf{Rationale.} Current schedulers observe resource requests but not whether a computation is still useful if it completes late; our Experiment 3 shows a highly-utilised schedule can still lose a large fraction of scientific value. PVR lets the federation compare mechanisms by preserved value, connecting allocation to the ``productive use'' and mission-priority questions DSIT raised. The principal risk is incentive compatibility --- self-declared urgency invites exaggeration --- so it must be tested under truthful and strategic reporting, hence a pilot rather than a deployment.\\[4pt]
\textbf{Timescale.} Year 2--3, 18 months. Depends on FAIRC-1 and FAIRC-2; should precede any federation-wide value-based pricing.\\[2pt]
\textbf{Resources.} $\sim$2.0 FTE (1.0 postdoctoral researcher; 1.0 senior RSE); $\sim$\pounds230,000. Aligns with the portal and mission-led weighting.
\end{recbox}

\begin{recbox}{FAIRC-5 --- Govern federated job mobility}{Could Have}
\textbf{Recommendation.} Before opening unmetered shared channels between clusters, put in place coordinated placement rules --- minimum-gain thresholds, receiving-system utilisation protection, and switching-cost ``tolls'' --- that allow a job to move only when the benefit is clear and the receiving system's local users are protected. Treat naive ``send each job wherever it looks fastest'' routing as a baseline to improve on, not a target.\\[4pt]
\textbf{Rationale.} Federation lets an overloaded system borrow spare or specialised capacity, but our Experiment 4 shows uncoordinated, individually-greedy placement behaves like selfish routing: it minimises global delay yet can harm the receiving system's local users and risk congestion under load spikes --- the computing analogue of Braess's paradox. Coordinated rules recover most of the benefit while limiting local harm and concentrating transfers in the genuine relief direction. We note honestly that naive currently looks best on aggregate metrics; the case for coordination rests on protecting local service and stability, and strengthening the local-harm evidence is part of the remaining work.\\[4pt]
\textbf{Timescale.} Year 3--4, 18 months. Long-term; depends on FAIRC-6 and FAIRC-1; runs with the phased expansion of the federated portfolio.\\[2pt]
\textbf{Resources.} $\sim$1.5 FTE (1.0 systems/network engineer; 0.5 economist/game theorist); $\sim$\pounds150,000. Aligns with EuroHPC/Waldur meta-scheduling.
\end{recbox}

% =================== CONCLUSION ===================
\section{Future Work and Research Directions}
\label{sec:future}

The recommendations in Section~\ref{sec:recs} are deliberately grounded in what can be done now with existing infrastructure. They also open onto a richer research agenda that FAIR-Compute is well placed to pursue, and which we believe the NFCS should be ambitious about, because the questions our interviews and simulations raised are exactly the ones a maturing federation will have to answer.

\paragraph{From heuristics to provable guarantees.} Experiment~2 shows tuned Slurm-like rules landing close to the offline benchmark, but only empirically, on particular windows. A natural next step is theoretical: establishing \emph{approximation guarantees} for transparent priority rules under stated assumptions on workload structure and contention. If simple, operator-maintainable heuristics can be proven to approximate welfare, fairness, or slowdown objectives, the federation gains a principled --- not merely observed --- licence to keep the schedulers it already runs.

\paragraph{Value-aware allocation and incentive compatibility.} The value experiment (Experiment~3) and recommendation FAIRC-3 point to a Projected Value Remaining mechanism, but the deep problem is \emph{incentive compatibility}: if priority depends on self-declared urgency or value, users will exaggerate. The research programme here is to design mechanisms --- potentially with artificial currencies, credits, or carefully bounded priority tokens --- under which truthful reporting of time-sensitivity is (approximately) optimal, and to characterise the efficiency--fairness frontier such mechanisms can reach. This connects directly to the classical literature on mechanism design without money and to dynamic assignment with limited manipulability.

\paragraph{Federation as a priced congestion game.} Experiment~4 treats federation through a congestion-game lens but stops at simple tolls. The fuller agenda is to quantify the \emph{price of anarchy} of uncoordinated cross-system routing on realistic federated topologies, and to design market-clearing tolls or shadow prices that provably realign self-interested placement with global welfare while protecting receiving systems. Doing this well requires the local-harm metrics we flag as outstanding, and it is where transport economics and algorithmic game theory have the most to offer the DRI.

\paragraph{Multi-resource clearing and dynamic boundaries.} JASMIN's storage-binding constraint (Section~\ref{sec:frictions}) motivates extending the single-resource model to a \emph{dual-resource} (compute-and-storage) clearing problem, and the batch-versus-interactive tension motivates treating the boundary between batch and software-defined environments as itself a dynamic allocation decision rather than a static partition. Both are concrete, high-value modelling problems surfaced directly by operators.

\paragraph{Learning, newcomers, and carbon.} Finally, several softer directions recur in the interviews: learning user ``types'' and workload profiles over time so that history-based mechanisms (FAIRC-8) treat newcomers fairly through sandbox or experience-based grace periods; rewarding accurate estimation with bonus credits; and making allocation \emph{carbon-aware}, reporting and eventually optimising tonnes of CO\textsubscript{2} equivalent alongside utilisation. Each is a modest extension individually, and collectively they describe a federation that not only allocates efficiently but learns, includes, and decarbonises as it scales.

Underpinning all of this is a single dependency we cannot over-state: \emph{data}. Every direction above becomes sharper, and testable, the moment the federation can supply real, pseudonymised job logs at scale. That is why FAIRC-1 is both the most mundane and the most enabling recommendation in this report, and why, given access to such data, we are eager to move the simulation evidence from sandbox to operational validation and to pursue the mechanism-design questions above in earnest.

\section{Conclusion}
\label{sec:conclusion}

FAIR-Compute set out to treat federated compute allocation as an economic and mechanism-design problem, not only an engineering one, and to test that lens against real practice and simulation. The consistent message is that the UK federation should \emph{measure first}: without a shared data standard and an efficiency KPI, neither operators nor funders can tell whether resources are used productively, and none of the more ambitious mechanisms can be evaluated. Beyond measurement, the near-term wins are unglamorous but real --- tune the transparent schedulers already in place, add a rapid-access tier, and standardise cross-system benchmarking --- while the genuinely novel research frontier is value-aware allocation and the careful governance of federated mobility, where uncoordinated ``selfish'' routing can quietly undermine the very efficiency federation is meant to deliver. We would welcome the opportunity to pressure-test these recommendations against DSIT and UKRI policy constraints and, given access to real UK job logs, to move the simulation evidence from sandbox to operational validation.

% =================== APPENDICES ===================
\appendix

\section{Strategic-reporting mechanisms}
\label{app:strategic}
This appendix develops the strategic-reporting part of the model (Section~\ref{sec:model-utility}). It is the mechanism-design foundation for recommendation FAIRC-8 and the future work of Section~\ref{sec:future}, and is \emph{not} exercised by the simulations, which implement only the descriptive scheduler and the offline benchmark.

\paragraph{Termination and requeue.} How an over- or under-estimate is resolved depends on the mechanism. Under a \emph{kill-at-walltime} rule the job is terminated the moment it exceeds its reservation, so its completion time is $C_j = s_j+l_j$ if $l_j\le l'_j$ and $C_j=\infty$ otherwise (an over-run yields no value). Under a gentler \emph{preempt-and-requeue} rule the job is not lost but delayed, incurring a requeue penalty $R_j$:
\begin{equation}
C_j = \begin{cases} s_j+l_j, & l_j\le l'_j,\\[2pt] s_j+l'_j+R_j+(l_j-l'_j), & l_j> l'_j. \end{cases}
\end{equation}

\paragraph{Penalty mechanisms.} The ex-post penalty $h_j(l'_j,l_j,H_{i(j)})$ can take three canonical forms, each mapping onto observed practice:
\begin{align}
U_j^{\mathrm{over}} &= \begin{cases} V_j-\kappa_jW_j-\alpha_{\mathrm{over}}(l'_j-l_j)^+, & l_j\le l'_j,\\ 0, & l_j>l'_j, \end{cases} \tag{overestimation charge}\\[4pt]
U_j^{\mathrm{fs}} &= \begin{cases} V_j-\kappa_jW_j-g_j(H_{i(j)}), & l_j\le l'_j,\\ -g_j(H_{i(j)}), & l_j>l'_j, \end{cases} \tag{fairshare adjustment}\\[4pt]
U_j^{\mathrm{prm}} &= \begin{cases} V_j-\kappa_jW_j, & l_j\le l'_j,\\ V_j-\kappa_j(W_j+R_j)-\alpha_{\mathrm{prm}}(l_j-l'_j)^+, & l_j>l'_j. \end{cases} \tag{preempt-and-requeue}
\end{align}
The overestimation charge prices the excess reservation directly; the fairshare adjustment routes the penalty through the evolving history state $H_{i(j)}=G(H_{i(j)},l'_j,l_j)$, so that accounting by \emph{reported} rather than actual usage lowers the user's future priority --- exactly the loop IRIS-SCRC and DiRAC operate; and preempt-and-requeue replaces the harsh kill with a bounded delay, partially insuring users against genuine uncertainty while preserving a deterrent against gross over-reporting. Characterising truthful, welfare-improving mechanisms of this kind, with and without artificial currency, is the mechanism-design programme flagged in Section~\ref{sec:future}.

\section{Evaluation metrics}
\label{app:metrics}
For completeness, we define the metrics used throughout Section~\ref{sec:sim}. For a job $j$ with submission time $a_j$, start $s_j$, completion $c_j$, and runtime $l_j$: the \emph{waiting time} is $W_j=s_j-a_j$; the \emph{turnaround} is $c_j-a_j$; and the \emph{slowdown} is $(c_j-a_j)/l_j$, which normalises delay by job length so that a short job made to wait is penalised more than a long one. \emph{Platform utilisation} is the fraction of available core-time actually used over the evaluation window; \emph{throughput} is jobs completed per unit time; \emph{makespan} is the time to clear the workload. For fairness we report, per account and per job-size class, the \emph{slowdown gap} (max$-$min average slowdown) and \emph{Jain's index} $J(x)=\left(\sum_i x_i\right)^2 / \left(n\sum_i x_i^2\right)$, applied both to allocated CPU service (a desirable quantity, where a higher index means more even sharing) and to delay (a cost, where a high index can also mean uniformly poor service). Because a high Jain index on delay is ambiguous, we always read it together with the average level of delay: the target is low average slowdown \emph{and} a high slowdown-Jain index.

\section{Inferred QoS-tier proxy}
\label{app:qos}
The public Fresco trace records account identifiers and per-job resources but not the true ACCESS project type. To give the simulator realistic account-scale heterogeneity we infer a proxy tier from observed demand: each account's monthly usage is annualised (CPU jobs measured in core-hours; whole-node jobs in node-equivalent core-hours at 128 cores/node; high-memory and GPU jobs charged with multipliers) and mapped to the four ACCESS-style tiers (Table~\ref{tab:qos}). The proxy should be read as an inferred demand tier, not the true administrative QoS; conclusions are not sensitive to the exact cut-points.

\begin{table}[htbp]\centering
\caption{ACCESS-style QoS-tier proxy inferred from annualised account demand.}
\label{tab:qos}
\small
\begin{tabularx}{\linewidth}{@{}llX@{}}
\toprule
\textbf{Proxy tier} & \textbf{Annualised usage proxy} & \textbf{Interpretation} \\
\midrule
Explore (1) & $\le 400{,}000$ & Small-scale or exploratory demand \\
Discover (2) & $(400{,}000,\,1{,}500{,}000]$ & Moderate demand \\
Accelerate (3) & $(1{,}500{,}000,\,3{,}000{,}000]$ & Larger mid-scale demand \\
Maximize (4) & $> 3{,}000{,}000$ & Highest inferred demand tier \\
\bottomrule
\end{tabularx}
\end{table}

\section{Federation policies and complete results}
\label{app:federation}
The federation broker (Experiment~4) is invoked at each job's arrival; it builds the feasible execution options, estimates each option's completion time from the unfinished work already on the target system, applies the active policy, and submits the job. The policies compared are defined in Table~\ref{tab:fed-policies}; the switching cost is a fixed component plus 10\% of the selected runtime, applied only to remote execution. Table~\ref{tab:fed-full} gives the complete results across all four workload scenarios, extending the selected rows shown in Section~\ref{sec:sim}.

\begin{table}[htbp]\centering
\caption{Federation placement policies (Experiment 4), in increasing order of coordination.}
\label{tab:fed-policies}
\small
\begin{tabularx}{\linewidth}{@{}lX@{}}
\toprule
\textbf{Policy} & \textbf{Rule} \\
\midrule
No federation & Jobs run only on their home system. \\
Naive & Place on the eligible system with the lowest estimated completion time. \\
Switching-cost & Choose the lowest estimated completion time \emph{plus} a remote switching toll (fixed $+\,10\%$ of runtime). \\
Minimum-gain & Move remotely only if the estimated saving exceeds a threshold $\theta$. \\
Utilisation-protected & Move only if minimum-gain holds \emph{and} the receiving system stays below a utilisation ceiling $U_{\max}$. \\
\bottomrule
\end{tabularx}
\end{table}

\begin{table}[htbp]\centering
\caption{Complete federation results across all workload scenarios (Experiment 4). ``Hom.''/``Het.''\ denote homogeneous (CPU-only) and heterogeneous (CPU+GPU) hardware.}
\label{tab:fed-full}
\small
\begin{tabularx}{\linewidth}{@{}llYYYY@{}}
\toprule
\textbf{Scenario} & \textbf{Policy} & \textbf{Avg wait} & \textbf{Avg slow.} & \textbf{$A\!\to\!B$} & \textbf{$B\!\to\!A$} \\
\midrule
\multirow{5}{*}{Asym., Hom.} & No federation & 127.8 & 13.33 & 0 & 0 \\
 & Naive & 4.0 & 1.39 & 5{,}974 & 13{,}402 \\
 & Switching cost & 18.9 & 2.17 & 739 & 22{,}466 \\
 & Minimum-gain & 4.7 & 1.61 & 1{,}748 & 8{,}316 \\
 & Utilisation-protected & 9.0 & 1.18 & 1{,}748 & 8{,}751 \\
\addlinespace
\multirow{5}{*}{Asym., Het.} & No federation & 9.5 & 1.60 & 0 & 0 \\
 & Naive & 0.2 & 1.00 & 6{,}822 & 9{,}108 \\
 & Switching cost & 8.2 & 1.06 & 1{,}371 & 20{,}408 \\
 & Minimum-gain & 0.3 & 1.00 & 2{,}801 & 7{,}080 \\
 & Utilisation-protected & 0.5 & 1.00 & 3{,}051 & 7{,}682 \\
\addlinespace
\multirow{5}{*}{Bal., Hom.} & No federation & 133.8 & 16.50 & 0 & 0 \\
 & Naive & 51.3 & 8.88 & 11{,}770 & 13{,}497 \\
 & Switching cost & 71.6 & 11.81 & 7{,}179 & 14{,}692 \\
 & Minimum-gain & 58.0 & 9.78 & 5{,}274 & 6{,}116 \\
 & Utilisation-protected & 69.6 & 10.20 & 3{,}117 & 4{,}435 \\
\addlinespace
\multirow{5}{*}{Bal., Het.} & No federation & 20.0 & 3.83 & 0 & 0 \\
 & Naive & 1.3 & 1.21 & 14{,}502 & 9{,}062 \\
 & Switching cost & 9.0 & 1.28 & 7{,}633 & 12{,}073 \\
 & Minimum-gain & 1.0 & 1.16 & 7{,}514 & 5{,}125 \\
 & Utilisation-protected & 3.8 & 1.50 & 7{,}454 & 5{,}075 \\
\bottomrule
\end{tabularx}
\end{table}

\section{Stakeholder engagement programme}
\label{app:engagement}
Table~\ref{tab:engagement} summarises the organisations engaged through interviews and follow-up during February--June 2026. Engagement was iterative: findings from earlier sessions shaped the agendas of later ones, and operational insights were mapped to the model variables and recommendations as described in Section~\ref{sec:method}.

\begin{table}[htbp]\centering
\caption{Interview and engagement programme (roles anonymised to organisation level).}
\label{tab:engagement}
\small
\begin{tabularx}{\linewidth}{@{}lX@{}}
\toprule
\textbf{Organisation} & \textbf{Focus of engagement} \\
\midrule
QMUL ITS Research (Apocrita) & Tier-3 operations, cost-recovery, fairshare, walltime behaviour \\
STFC IRIS / IRIS-SCRC & RSAP allocation cycle, ex-post penalty loops, computing-model bidding \\
JASMIN (NERC) & Data-intensive allocation, storage governance, residual-claimant federation \\
DiRAC & Credit-based allocation, ``use-it-or-lose-it'', under-utilisation review \\
EPCC / NFCS coordination & National service transition, federation coordination, reporting \\
UKRI Digital Research Infrastructure & Roadmap, ROI/KPIs, disaggregated stack, CCEs \\
DSIT public-compute division & AIRR allocation, ``productive use'', policy objectives \\
\bottomrule
\end{tabularx}
\end{table}

\section{Software and reproducibility}
\label{app:software}
The simulation framework is implemented in Python. Multi-objective priority-weight search uses \textbf{pymoo} with the NSGA-II algorithm~\citep{blank2020pymoo,pymoo2026,nsga2002}; the offline welfare benchmark is a constraint-programming model solved with \textbf{PyJobShop}~\citep{lan2025pyjobshop,pyjobshopdocs2026}; workload-cluster interpretability uses \textbf{SHAP}~\citep{shap_tools}. Single-system experiments use a representative 7{,}000-job high-pressure window on a simulated platform of 300 nodes $\times$ 128 cores (38{,}400 CPU cores); the deadline-value study uses a 500-job window to keep the offline optimisation tractable; the federation study uses two such systems with 30-day synthetic traces. All contention results are point estimates from these windows and are labelled as synthetic sandbox experiments throughout. The event-driven federation simulator underlying Experiment~4 --- covering homogeneous and heterogeneous (CPU/GPU) two-system federation, the workload scenarios, switching costs, and the local-user-protection rules --- is openly available at \url{https://github.com/angelamath/HPC_federation_simulator}.

% =================== TEAM & ACK ===================
\section*{Team and Acknowledgements}
\addcontentsline{toc}{section}{Team and Acknowledgements}
FAIR-Compute is led by Professor Konstantinos (Kostas) E.\ Zachariadis (Principal Investigator, School of Economics and Finance, Queen Mary University of London), with Dr Ahmed Sayed (co-lead, Electronic Engineering and Computer Science, QMUL) and Professor Dimitris Fotakis (advisor, Electrical and Computer Engineering, National Technical University of Athens). Angeliki Mathioudaki (NTUA and ICCS) led the simulation study and model; Wan Shuen Siaw (School of Mathematical Sciences, QMUL) led the landscape review and field data. We thank the many operators and policymakers who gave their time in interviews and the stakeholder survey, including colleagues at QMUL ITS Research, IRIS, JASMIN, DiRAC, EPCC, DSIT, and the UKRI Digital Research Infrastructure programme. We acknowledge the support of the National Federated Compute Services NetworkPlus, who have funded this work through the Engineering and Physical Sciences Research Council (EPSRC) under Grant No.\ EP/Z534493/1 (NFCS Network+ Flexible Fund).

% =================== REFERENCES ===================
\bibliographystyle{plainnat}
\bibliography{references}
\addcontentsline{toc}{section}{References}

\end{document}